\documentclass[aps,pra,reprint,twocolumn,showpacs,showkeys,floatfix,nobibnotes]{revtex4-2}

\usepackage{hyperref}
\hypersetup{colorlinks=true, urlcolor=red, citecolor=blue, linkcolor=blue}
\usepackage{graphicx}
\usepackage{subfigure}
\usepackage{epstopdf}
\usepackage{lastpage,physics}
\usepackage{braket}

\begin{document}
\title{Nonlinear displaced Kerr state and its nonclassical properties}

\author{Arpita Chatterjee}
\affiliation{School of Physical Sciences, Jawaharlal Nehru University, New Delhi 110067, India}
\author{Rupamanjari Ghosh}
\affiliation{School of Physical Sciences, Jawaharlal Nehru University, New Delhi 110067, India}
\affiliation{Shiv Nadar University, Gautam Buddha Nagar, UP 201314, India}

\email{Corresponding author: rghosh.jnu@gmail.com}

\date{Compiled \today}



\begin{abstract}
We construct a distinct class of nonlinear displaced Kerr state by application of the displacement operator upon a state which is prepared by sending the well-known photon-added coherent state through a normal Kerr medium. A sketch for the experimental set-up for preparing the state is suggested. We evaluate some statistical properties such as the photon number distribution, Mandel's $Q$ parameter, Husimi-Q and Wigner functions, and quadrature squeezing, for the nonlinear displaced Kerr state, and then analyze the nonclassicality in terms of these standard parameters. We reduce the infinite-level problem to a truncated discrete two-level system by using low Kerr parameter approximation and then convert the generated nonclassicality into bipartite entanglement between the two modes of an output state of a linear optical device.
\end{abstract}



\maketitle


\section{Introduction}
\label{sec1}

The term ``coherent states'' was introduced by Glauber \cite{glauber63} in the context of quantum optics to characterize those states of the electromagnetic field that factorize the field coherence function to all orders. Glauber constructed the field coherent states by using the harmonic oscillator algebra and concluded that the states can be derived from any one of the three following mathematical definitions: i) as eigenstates of the annihilation operator $\hat{a}|\alpha\rangle=\alpha|\alpha\rangle$; ii) as states obtained by the application of the displacement operator upon the vacuum state $|\alpha\rangle=\hat{D}(\alpha)|0\rangle$ with $\hat{D}(\alpha)=e^{\alpha\hat{a}^\dag-\alpha^*\hat{a}}$ and iii) as
the quantum states with a minimum uncertainty relationship $(\triangle\hat{p})^2(\triangle\hat{q})^2=(1/2)^2$, where the coordinate and momentum operators are defined as $\hat{q}=(1/\sqrt{2})(\hat{a}+\hat{a}^\dag)$, $\hat{p}=(i/\sqrt{2})(\hat{a}^\dag-\hat{a})$ with the condition $\triangle\hat{p}=\triangle\hat{q}=1/\sqrt{2}$ \cite{glauber163,puri96}.

Agarwal and Tara \cite{agarwal91} proposed, theoretically, a new class of non-Gaussian states, which is intermediate between the coherent state $|\alpha\rangle$ (the most classical-like quantum state) and the number state $|n\rangle$ (purely quantum state), by repeated application of the photon creation operator on the coherent state basis. The resulting class of states was identified as nonclassical states which are promising candidates for applications in quantum information technology \cite{tittel98}. Almost after a decade, Zavatta \textit{et al.} \cite{zavatta04} reported a single photon excitation of a classical coherent field experimentally, and an ultra-fast, time-domain, quantum homodyne tomography technique was used to describe a quantum to classical transition. The degaussification process can be realized in a simple manner by adding (subtracting) photons to (from) a Gaussian field, and the resulting states are known to exhibit nonclassical properties such as negativity of the Wigner function \cite{hillery84}, antibunching \cite{kimble77}, sub-Poissonian photon statistics \cite{short83} or squeezing in one of the quadratures of the field \cite{dodonov02}, etc. For example, a traveling non-Gaussian field was produced by subtracting a photon from a squeezed vacuum \cite{wenger04} in another experiment. But while the pulsed homodyne detection scheme confirmed non-Gaussian statistics for the photon subtracted squeezed vacuum, the Wigner function reconstructed from the experimental data failed to exhibit negativity.  Kim \textit{et al.} \cite{kim05} explained that unless the input Gaussian radiation is nonclassical, one cannot generate a nonclassical field through photon subtraction. They also pointed out the contrasting situation of photon addition to Gaussian fields, where even a highly classical state like the thermal state turns out to be nonclassical \cite{lee95,jones97}. The rapidly developing area of quantum computation and information theory has kindled further interest in generating and manipulating nonclassical radiation fields in continuous variable quantum states. In such a context, it is timely to analyze the changes in the nonclassical features of a displaced Kerr state due to photon addition.

The nonlinear coherent states or $f$-coherent states $|\alpha\rangle_f$ have been introduced \cite{filho96,manko97} as eigenstates of a deformed annihilation operator $\hat{A}|\alpha\rangle_f=\alpha|\alpha\rangle_f$, where $\hat{A}=\hat{a}f(\hat{N})$ with $f(\hat{N})$ a deformation function of the number operator
$\hat{N}=\hat{a}^\dag\hat{a}$, and also by the application of a deformed displacement operator $\hat{D}_{D}(\alpha)=\exp{(\alpha\hat{A}^\dag-\alpha^*\hat{A})}$ upon the vacuum state, i.e., $|\alpha\rangle_{D}=\hat{D}_{D}(\alpha)|0\rangle$. These two operations correspond to the generalization of Glauber's first two definitions for the construction of a coherent state. These types of states also show nonclassical properties \cite{recamier06}. Roman \textit{et al.} \cite{roman14} constructed a deformed photon-added nonlinear coherent state (DPANCS) by applying the deformed creation operator upon the nonlinear coherent state, obtained by using two different generalizations: one is as the eigenstate of the deformed annihilation operator, and the other is by applying a deformed displacement operator upon the vacuum state. They evaluated some statistical properties such as Mandel's $Q$ parameter, Husimi and Wigner functions, and found a profound difference in these statistical properties of the DPANCSs obtained from the two different generalizations.

\begin{figure*}[ht]
\centering
\includegraphics[width=14cm]{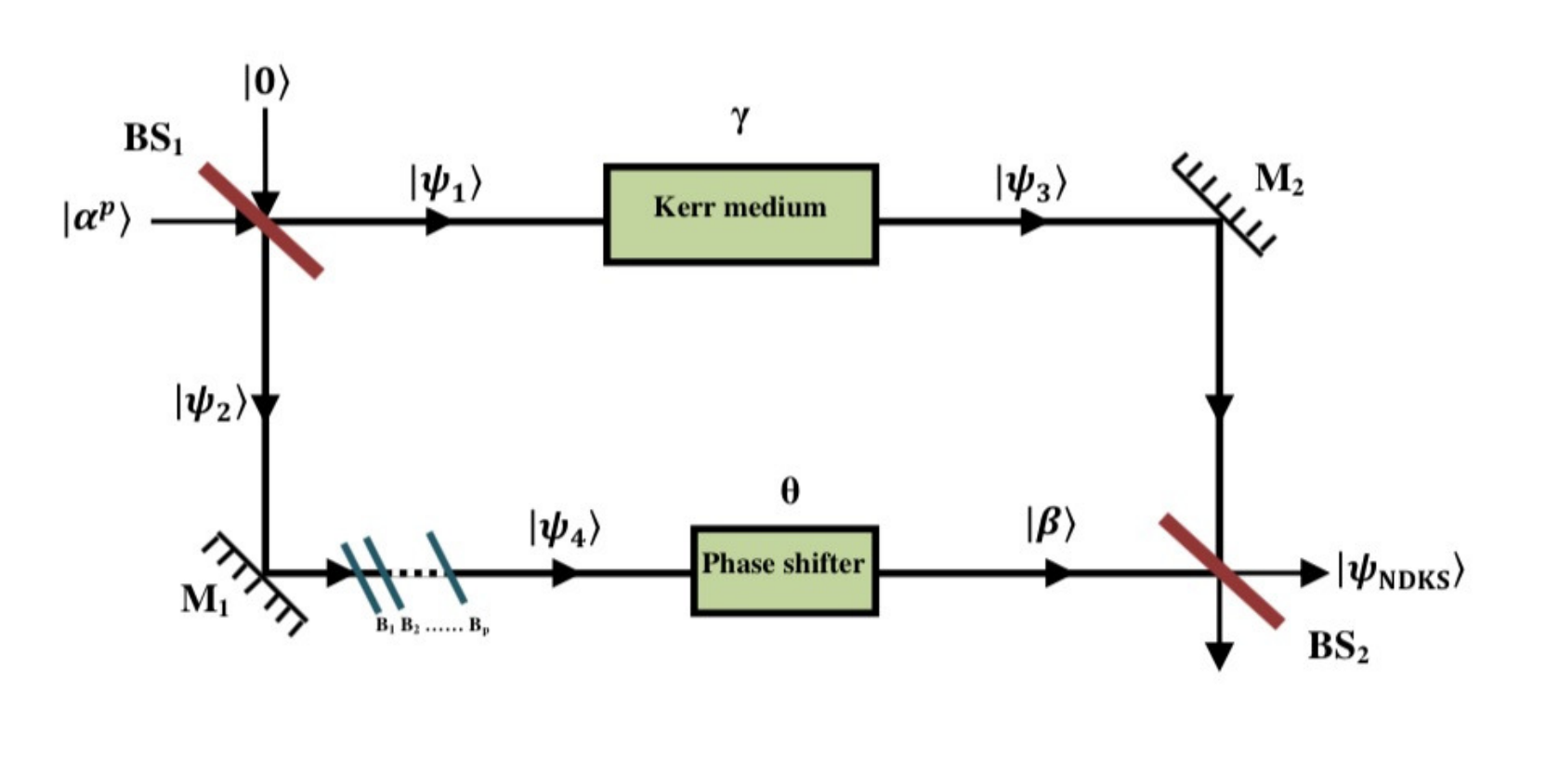}

\caption{(Color online) Sketch of the proposed experiment for preparing a nonlinear displaced Kerr state}
\label{fig1}
\end{figure*}

We begin by introducing an experimental set-up for preparing a nonlinear displaced Kerr state (NDKS). The apparatus used to generate and analyze the NDKS is schematically described in Fig.~\ref{fig1}. It is a nonlinear Mach-Zehnder interferometer containing an optical Kerr medium in one arm and a phase-shifter in another arm \cite{shiraski90,shiraski91}. If the coupling of the optical field with reservoirs such as optical-loss oscillators and the atomic system of the Kerr medium are adiabatically eliminated, this interferometer can be treated as a quantum system with two input ports and two output ports. The first beam-splitter $\mathrm{BS}_1$ is used to simply superpose a $p$-photon-added-coherent-state input signal with an input vacuum field. The state entering into the Kerr medium in the upper arm of the interferometer is a $p$-photon-added-coherent state, modified by the transmissivity of the first beam-splitter and the associated vacuum state of the second input port. The nonclassical light after propagating through a `distance' of $\gamma$ in the Kerr medium is expressed as a nonlinear Kerr state. The other input field, after reflecting from the mirror $M_1$, proceeds through $p$ numbers of consecutive high-transmissivity beam-splitters. These beam-splitters $\mathrm{B}_1,~\mathrm{B}_2,~...~,\mathrm{B}_p$ are used to subtract $p$ photons and thus to transfer the input field to a simple coherent state. The state $|\psi_4\rangle$ remains a coherent state $|\psi_4 e^{i\theta}\rangle$ after crossing the phase shifter. The second beam-splitter $\mathrm{BS}_2$ is used to merge the nonlinear Kerr state in the upper arm with the coherent state in the lower arm of the device, and this combination acts to displace the nonlinear Kerr state and give a final output state. The high reflectivity of $\mathrm{BS}_2$ prevents the output state from being contaminated by the state $|\psi_4 e^{i\theta}\rangle$.

The manufacturing of the NDKS can be shown mathematically by using standard operator notations as the following. When a $p$-photon-added coherent state $|\alpha^p\rangle$ \cite{agarwal91} is incident upon a beam-splitter $\mathrm{BS}_1$ with the other input field in a vacuum state, two $p$-photon-added coherent states are prepared behind $\mathrm{BS}_1$ as \cite{kitagawa86}
\begin{eqnarray}\nonumber
|\psi_1\rangle & = & |\alpha_1^p\rangle,~~~\alpha_1^p=(1-r_1)^{1/2}\alpha^p,\\\nonumber
|\psi_2\rangle & = & |\alpha_2^p\rangle,~~~\alpha_2^p=r_1^{1/2}\alpha^p.
\end{eqnarray}
Here $r_1$ is the reflectivity of $\mathrm{BS}_1$, and $\alpha_1^p$ and $\alpha_2^p$ are c-numbers. A nonlinear Kerr medium interacting with the state $|\psi_1\rangle$ generates the appropriate intensity-dependent phase shift through an anharmonic coupling described by the Hamiltonian \cite{yurke86,gerry94,wilson91}
\begin{equation}
H = \hbar\hat{a}^\dag \hat{a}+\hbar\chi\hat{a}^{\dag2}\hat{a}^2,
\end{equation}
where the coupling coefficient $\chi$ is proportional to the third-order nonlinear susceptibility of the Kerr medium. The output state from the Kerr medium is
\begin{equation}
|\psi_3\rangle = \hat{U}_K(\gamma)|\psi_1\rangle,
\end{equation}
where $\hat{U}_K(\gamma)$ is the evolution operator in the interaction picture
\begin{equation}
\hat{U}_K(\gamma) = \exp[i\frac{\gamma}{2}\hat{a}^{\dag2}\hat{a}^2]=\exp[i\frac{\gamma}{2}\hat{N}(\hat{N}-1)],
\end{equation}
with $\gamma \equiv 2\chi L/v$, $\chi$ being related to the Kerr medium third-order susceptibility, $L$ the length of the medium, and $v$ the appropriate phase velocity inside the medium. The high-transmissivity beam-splitters $\mathrm{B}_1,~\mathrm{B}_2,~...~,\mathrm{B}_p$, situated one after another at the lower arm of the device, are used for pulling out $p$ photons from the state $|\psi_2\rangle$ \cite{zavatta08,chatterjee12,benlloch12}. After passing through these $p$-number of beam-splitters, $|\psi_2\rangle$ changes to a simple coherent state $|\psi_4\rangle$. Then the phase modulator just applies a phase shift of $\theta$ to the state $|\psi_4\rangle$, and it remains a coherent state $|\psi_4 e^{i\theta}\rangle (|\beta\rangle,~\mathrm{say})$ \cite{pinheiro13}. If the next beam-splitter $\mathrm{BS}_2$ has sufficiently high reflectivity $r_2\sim 1$, then quantum fluctuations from the lower arm are not mixed into the output port 1 at $\mathrm{BS}_2$. Therefore, the contribution of $|\psi_4 e^{i\theta}\rangle$ to the output state is merely a classical driving force, which can be described by a unitary displacement operator $\hat{D}(\beta)$ \cite{kitagawa86}. Finally, the output field at port 1 of $\mathrm{BS}_2$ is written as
\begin{eqnarray}
|\psi_{\mathrm{NDKS}}\rangle = \hat{D}(\beta)|\psi_3\rangle =  \hat{D}(\beta)\hat{U}_K(\gamma)|\alpha_1^p\rangle.
\end{eqnarray}
Here $\hat{U}_K(\gamma)$ is a unitary operator connecting the input $|\psi_1\rangle$ and the output $|\psi_3\rangle$ of the Kerr medium, i.e., representing the self-phase modulation. The interference at the beam-splitter $\mathrm{BS}_2$ is represented by the displacement operator $\hat{D}(\beta)$. In this way, a nonlinear Mach-Zehender interferometer blends a nonlinear Kerr transformation with a displacement in phase-space.

It is to be noted that here we treat photon subtraction as an ideal $\hat{a}$ operation. In realistic schemes based on the beam-splitter, this approximation becomes exact only in the nonphysical limit of vanishing interaction (e.g., for perfect transmissivity of the beam splitter). However, as long as the interaction is kept very weak, which makes successful subtraction a rare event but still frequent enough for applications, the idealized description is a good approximation \cite{kim08}. Conventional materials are known to offer only a very small Kerr nonlinear response. The ease of practical implementation of our scheme will thus depend on `preparing’ a medium with large third-order susceptibility. There do exist a few situations, with dipole resonance conditions and/or cavity field confinement, in which large Kerr nonlinearities have been shown to be created even by a small number of photons \cite{sanders92,waks06}.

This paper is organized as follows. Different nonclassicality indicators used in our study are defined in Sec.~\ref{sec2}. In Sec.~\ref{sec3}, the basic equation describing the considered state is followed by the results obtained in terms of the photon number distribution, Mandel's $Q$ parameter, squeezing parameter and quasiprobability distribution functions like Wigner and Husimi-$Q$. Entanglement, another striking quantum feature, is of great importance in the development of quantum information processing \cite{tittel98}. There has been considerable progress in understanding the connection between nonclassicality and entanglement. It has been identified \cite{asboth05,paris99} that nonclassicality is an obvious source of entanglement. A beam-splitter is capable of converting nonclassicality of a single-mode radiation into bipartite entanglement. This property, viz. \textit{nonclassicality as an entanglement resource}, has been employed to identify the entanglement potential (EP) of single-mode radiation fields. EP allows us to analyze the degree of nonclassicality of a given single-mode radiation field. We compute EP of nonlinear displaced Kerr states and find that the EP increases with the increase of seed beam intensity, which is in confirmation with the analysis of the Wigner function of the state. Section~\ref{sec4} carries our conclusions.

\section{Nonclassicality measures}
\label{sec2}

Generally a quantum state is recognized as a nonclassical one if it cannot be expressed as a statistical mixture of coherent
states \cite{ushadevi06}. It has been well accepted that the non-existence of a well defined Glauber-Sudarshan $P$-function \cite{sudarshan63} implies nonclassicality of a given state. However, the evaluation of $P$-function introduces operational difficulties, as it requires complete information of the state to be examined. Various types of indicators, which can be used to distinguish between classical and nonclassical states in experimental measurements, have been proposed from the early days of quantum optics. Such signatures of nonclassicality, verifiable in simple experiments, are negative Wigner function \cite{hillery84}, antibunching \cite{kimble77}, sub-Poissonian photon statistics \cite{short83}, squeezing \cite{dodonov02}, photon number oscillations \cite{schliech87} etc. In this paper, we analyze several such measures to quantify the nonclassicality of a proposed quantum state.

One of the simplest quantitative ``measures of nonclassicality'' is the Mandel's $Q$ parameter. It quantifies the deviation from the Poissonian statistics. If $Q=0$, as in the case of a coherent state, the photon distribution is Poissonian. If $Q<0$, the state is undoubtedly nonclassical, having the sub-Poissonian photon distribution. Several well-known sets of nonclassical states are sub-Poissonian for all values of their parameters. For example, the set of binomial states $|p, M\rangle = \sum_{n=0}^M B_n^M|n\rangle$ with $B_n^M = \left[\binom{M}{n}p^n(1-p)^{M-n}\right]^{1/2}$ has $Q=-p$ for any value of $M$. When $Q>0$, we have a super-Poissonian field but one cannot address the state as ``classical''. For example, for all even coherent states, $Q>0$ but their Wigner functions take negative values in some part of the phase space.

For a single-mode system, characterized by the density operator $\hat{\rho}$, the Wigner function is defined through \cite{hillery84}
\begin{eqnarray}\nonumber
W(\zeta, \zeta^*) =  \frac{2}{\pi^2}e^{2|\zeta|^2}
\int{\langle-\alpha|\hat{\rho}|\alpha\rangle e^{-2(\zeta^*\alpha-\zeta\alpha^*)}d^2\alpha},
\end{eqnarray}
where $|\alpha\rangle$ is a coherent state. Basically the Wigner function is a quasiprobability distribution representing quantum states in phase space. It is not a true probability distribution as it can take negative values. If for a state, Wigner function takes any negative value, the quantum state has no classical analog. However, the converse does not hold good: when the Wigner function is positive everywhere, one cannot conclude that the state is classical. For example, for a squeezed state,
Wigner function is a Gaussian distribution and is positive throughout, but squeezed radiation is a well-known nonclassical field. One has to then resort to other measures of nonclassicality.

A coherent state is a minimum-uncertainty state, whose quadrature components behave symmetrically, and the fluctuations in the amplitudes of the two quadrature phases are the minimum. A state is defined as a quadruture-squeezed state only when either of its quadratures has a standard deviation that falls below the coherent state value. The uncertainty in one quadrature may be squeezed below the coherent-state value if the uncertainty in the other quadrature gets stretched so that the product obeys the Heisenberg uncertainty principle. A quadrature-squeezed state may, but need not, be a minimum-uncertainty state. In this paper, we discuss the squeezing property of the NDKS by using the standard analysis of quadrature squeezing. Quadrature squeezing implies the nonclassicality of the underlying field but sometimes this measure also fails to test the property. For example, a field given by the density matrix
\begin{eqnarray}\nonumber
\hat{\rho} = \hat{a}^{\dag m}e^{-\beta\hat{a}^\dag\hat{a}}\hat{a}^m,~~\beta>0,~m~\mathrm{integer}
\end{eqnarray}
exhibits no squeezing but is proved to be nonclassical in terms of some other criteria \cite{agarwal92}. Again, when a general superposition of the product (SUP) operations, $s\hat{a}\hat{a}^\dag + t\hat{a}^\dag \hat{a}$ is applied on a coherent state, where the scalars $s$ and $t$ are related by $s=\sqrt{1-t^2}$, the nonclassical feature is highlighted by its Wigner function distribution, but the resulted field does not show any squeezing property \cite{chatterjee12}.

In spite of the fact that the Husimi-$Q$ function is not related with the nonclassicality of a given field, sometimes it is convenient to use it for simple representations of quantum states. This distribution is always positive and is defined as the Fourier transform of the anti-normally-ordered characteristic function $\chi_{\mathrm{an}}(\alpha)$ as
\begin{eqnarray}\nonumber
Q(\xi) = \frac{1}{\pi^2}\int{d^2\alpha~e^{(\xi\alpha^*-\alpha\xi^*)}\chi_{\mathrm{an}}(\alpha)},
\end{eqnarray}
which is simply related to the expectation value of the density operator $\hat{\rho}$ in a coherent state $|\xi\rangle$ by
\begin{eqnarray}
Q(\xi) = \frac{1}{\pi}\langle\xi|\hat{\rho}|\xi\rangle.
\label{eq5}
\end{eqnarray}
Mundarain \textit{et al.} \cite{mundarain04} proposed a phase-space description of the photon number distribution of nonclassical states, based on the Husimi-$Q$ function. They illustrated this approach for displaced number states and two photon coherent states, and proved it to be an efficient method for computing and interpreting the photon number distribution.

We also investigate entanglement potential \cite{asboth05}, a quantum information-theoretic measure of nonclassicality, for our discussed nonlinear state. EP gives the amount of two-mode entanglement that can be originated by putting a given single-mode state in port $a$ and a vacuum state in port $b$ of a 50:50 beam-splitter and then mixing them. It is important to note that a classical single-mode radiation field does not get entangled in such an arrangement \cite{paris99}. Thus the entanglement produced in a linear optical device certifies the nonclassicality of the field entering into the same.

More specifically, the entanglement potential is calculated by using the logarithmic negativity (LN) \cite{vidal02} of a bipartite quantum state as
\begin{eqnarray}\nonumber
\mathrm{EP} = \mathrm{log}_2||\hat{\rho}_{ab}^{T_a}||_1,
\end{eqnarray}
where $\hat{\rho}_{ab}^{T_a}$ denotes the partial transpose of a two-mode density operator $\hat{\rho}_{ab} = \hat{\mathcal{B}}_{ab}(\hat{\rho}\otimes|0\rangle
\langle 0|)\hat{\mathcal{B}}^\dag_{ab}$ with respect to mode $a$. The symbol $||.||_1$ denotes the trace norm and $\hat{\mathcal{B}}_{ab}$ corresponds to a beam-splitter operator, i.e.
$\hat{\mathcal{B}}_{ab}=\exp\big[\frac{\pi}{2}(\hat{a}^\dag\hat{b}-\hat{a}\hat{b}^\dag)\big]$, whose action on the creation operators $\hat{a}^\dag$, $\hat{b}^\dag$ of the two input ports is explicitly given by
\begin{eqnarray}\nonumber
\hat{\mathcal{B}}_{ab}\hat{a}^\dag\hat{\mathcal{B}}^\dag_{ab} & = & (\hat{a}^\dag+\hat{b}^\dag)/\sqrt{2}\\\nonumber
\hat{\mathcal{B}}_{ab}\hat{b}^\dag\hat{\mathcal{B}}^\dag_{ab} & = & (\hat{a}^\dag-\hat{b}^\dag)/\sqrt{2}.
\end{eqnarray}
The LN is defined via the Peres-Horodecki criterion for positivity under partial transpose \cite{peres96,horodecki97}, which is a necessary condition for checking if a given state is separable or not. The condition is sufficient only for low-dimensional bipartite systems (qubit-qubit and qubit-qutrit) but ceases to be so in higher dimensions. EP has already been evaluated for a variety of nonclassical states such as squeezed states, even and odd coherent states, Fock states, etc. \cite{asboth05}. Here we use the LN as a measure of the bimodal entanglement to evaluate EP for the NDKS.

\section{The state and its properties}
\label{sec3}

A nonlinear coherent state \cite{filho96} or $f$-coherent state \cite{manko97} is introduced as the right eigenstate of a deformed annihilation operator $\hat{A}$, which is a product of the boson annihilation operator $\hat{a}$ and a nonlinear function $f(\hat{N})$ of the number operator $\hat{N}$,
\begin{equation}
\hat{A}|\alpha\rangle_f = \hat{a}f(\hat{N})|\alpha\rangle_f = \alpha|\alpha\rangle_f,
\end{equation}
where $\alpha$ is the complex eigenvalue. The well-known $p$-photon-added coherent state (PACS) \cite{agarwal91}
\begin{eqnarray}\nonumber
|\alpha^p\rangle = \frac{{\hat{a}^{\dag p}}|\alpha\rangle}{{[p!L_p(-|\alpha|^2)]}^{1/2}},
\end{eqnarray}
where $L_p(x)$ is the Laguerre polynomial of order $p$, is a nonlinear coherent state as it satisfies
\cite{sivakumar99}
\begin{equation}
\hat{a}\left(1-\frac{p}{1+\hat{a}^\dag \hat{a}}\right)|\alpha^p\rangle = \alpha|\alpha^p\rangle.
\end{equation}
A PACS of this type shows nonclassical properties like squeezing and sub-Poissonian statistics, and it can be produced in laser-atom interactions under appropriate conditions \cite{agarwal91}. Zavatta \textit{et al.} also proposed an experimental set-up for preparing a single-photon-added coherent state \cite{zavatta05}. Unlike the operation of photon annihilation, which maps a coherent state into another coherent state ($\hat{a}|\alpha\rangle=\alpha|\alpha\rangle$), i.e., a classical field into another classical field, the single-photon excitation of a coherent state changes it into something quite different, especially for low values of $\alpha$, where the absence of the vacuum term has a stronger impact. In the extreme case of an initial vacuum state $|0\rangle$, the addition of one photon indeed transforms it into the very nonclassical single-photon Fock state $|1\rangle$, which exhibits negative values of the Wigner function around the origin. Also the nonlinear coherent states are demonstrated to be useful tools for the description of the center-of-mass motion of a trapped ion \cite{filho96}.

Inspired by this, we wish to extend the concept of displaced Kerr state \cite{wilson91} to a new nonlinear one. The nonlinearity comes by replacing the canonical coherent input state to the Kerr medium by a nonlinear PACS. The resultant nonlinear displaced Kerr state is given by
\begin{equation}
|\psi_{\mathrm{NDKS}}\rangle = \frac{1}{N_1}\hat{D}(\beta)\hat{U}_K(\gamma)|\alpha^p\rangle,
\end{equation}
where $N_1$ is the normalization constant. It can be written explicitly in the number state basis as
\begin{eqnarray}\nonumber
\label{eq9}
|\psi_{\mathrm{NDKS}}\rangle & = & \frac{e^{-|\alpha|^2/2}}{[p!L_p(-|\alpha|^2)]^{1/2}}
\sum_{m=0}^\infty e^{i\frac{\gamma}{2}(m+p)(m+p-1)}\frac{\alpha^m}{m!}\\
& & \times\sqrt{(m+p)!}\hat{D}(\beta)|m+p\rangle.
\end{eqnarray}
We now analyze the statistical properties of the state constructed above.

\subsection{Photon number distribution}

The photon number distribution, viz. the probability of finding $n$ photons, is a key characteristic of every quantum state. Using \cite{tanas92}
\begin{eqnarray}\nonumber
& & \langle n|\hat{D}(\beta)|m\rangle\\
& = &
\left\{
\begin{array}{lcl}
e^{-|\beta|^2/2}\beta^{n-m}(\frac{m!}{n!})^{1/2} L_m^{n-m}(|\beta|^2),~n\geq m\\\\
e^{-|\beta|^2/2}(-\beta^*)^{m-n}(\frac{n!}{m!})^{1/2} L_n^{m-n}(|\beta|^2),~n<m,
\end{array}
\right.
\end{eqnarray}
where $L_n^m(x)$ is the associated Laguerre polynomial, the photon-number distribution $P_\textrm{NDKS}$ of Eq.~(\ref{eq9}) is obtained as
\begin{eqnarray}\nonumber
& & P_\textrm{NDKS}\\\nonumber
& = & |\langle n|\psi_{\mathrm{NDKS}}\rangle|^2\\
& = &
\left\{
\begin{array}{lcl}
\frac{e^{-|\alpha|^2-|\beta|^2}}{p!L_p(-|\alpha|^2)n!}\left|
\sum_{m=0}^\infty\frac{\alpha^m}{m!}(m+p)!e^{i\frac{\gamma}{2}(m+p)(m+p-1)}\right.\\
\left.\times\beta^{n-m-p}L_{m+p}^{n-m-p}(|\beta|^2)\right|^2,~n\geq m+p\\\\
\frac{e^{-|\alpha|^2-|\beta|^2}n!}{p!L_p(-|\alpha|^2)}\left|
\sum_{m=0}^\infty\frac{\alpha^m}{m!}e^{i\frac{\gamma}{2}(m+p)(m+p-1)}\right.\\
\left.\times(-\beta^*)^{m+p-n}L_{n}^{m+p-n}(|\beta|^2)\right|^2,~n<m+p.
\end{array}
\right.
\end{eqnarray}

\begin{figure*}[ht]
\centering
\includegraphics[width=8cm]{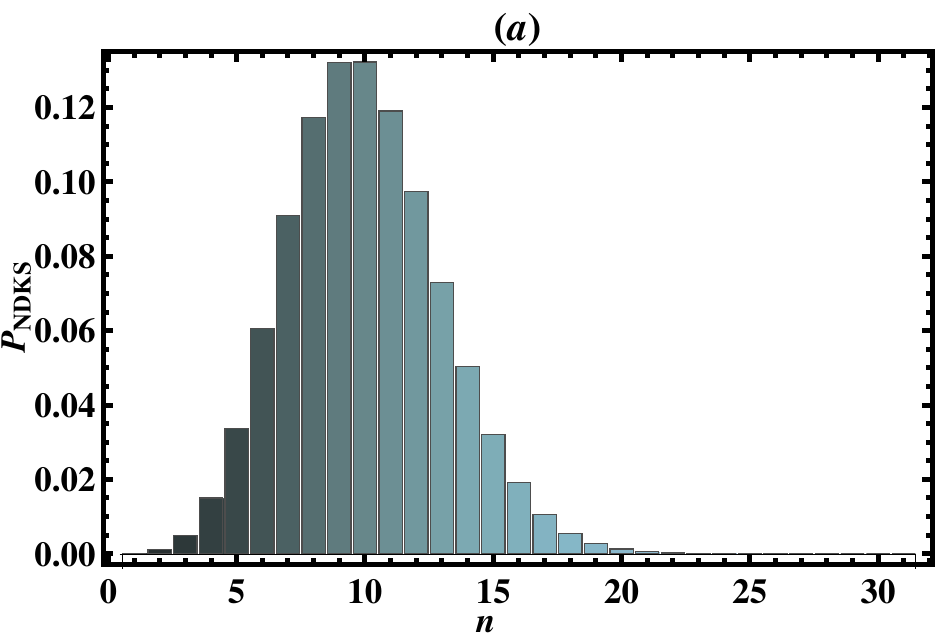}\hspace{1cm}
\includegraphics[width=8cm]{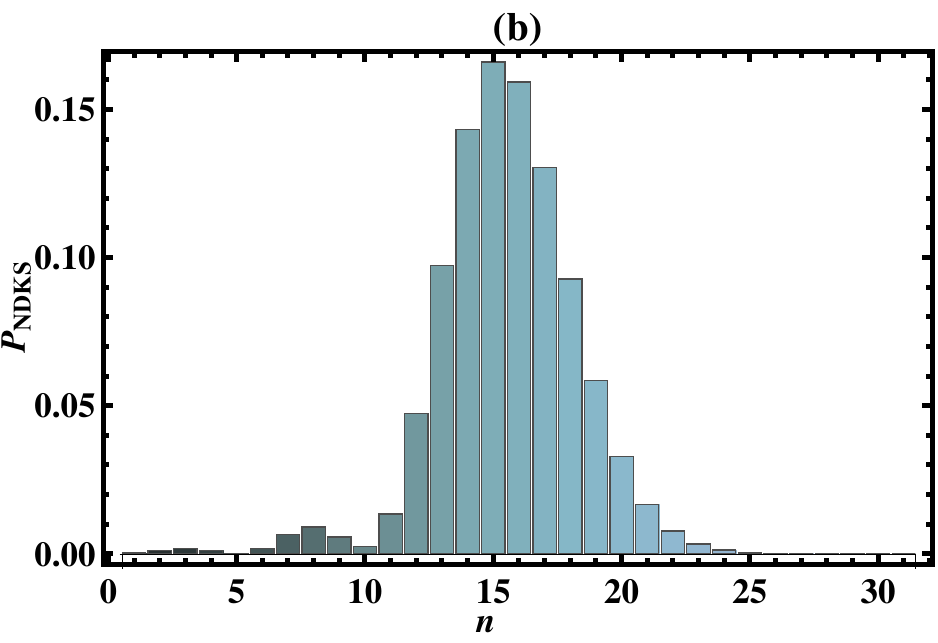}\hspace{1cm}\\
\includegraphics[width=8cm]{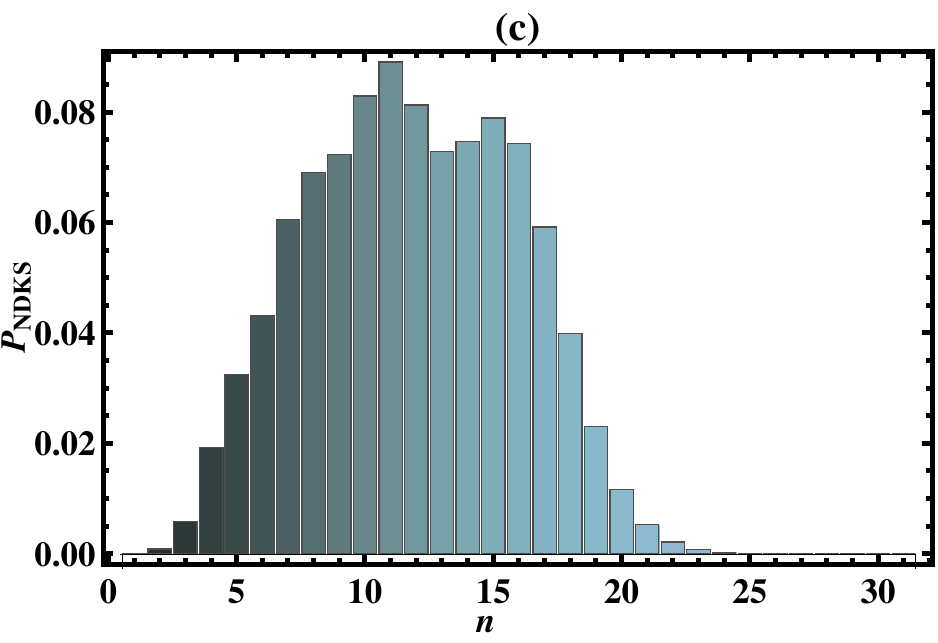}\hspace{1cm}
\includegraphics[width=8cm]{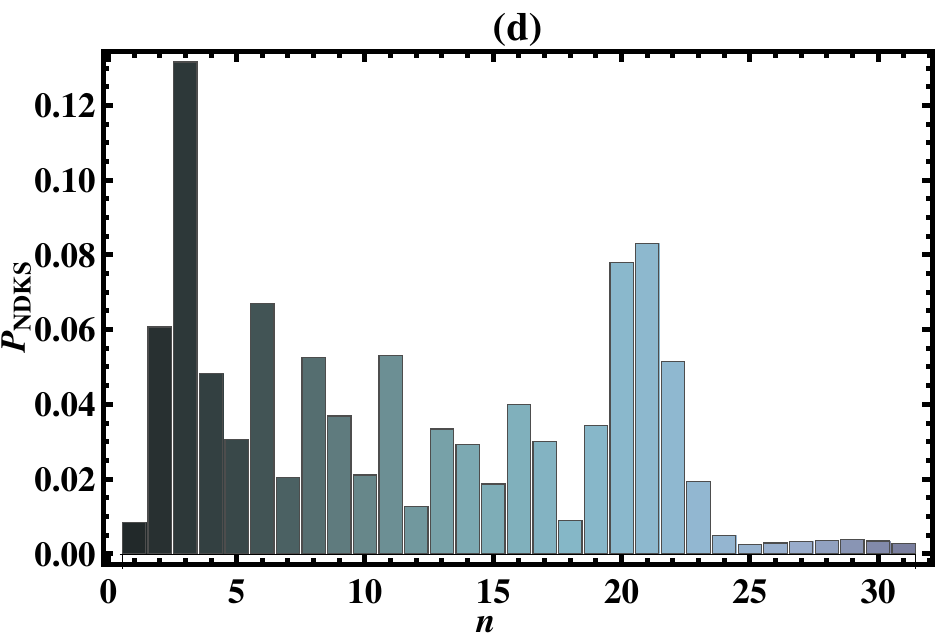}

\caption{(Color online) Photon-number distribution for the nonlinear displaced Kerr state with $(a)~\gamma=0.05,~p=0$; $(b)~\gamma=0.12,~p=2$; $(c)~ \gamma=0.25,~p=4$ and $(d)~\gamma=1,~p=6$. $\alpha$ and $\beta$ are taken to be real as $\alpha=1$ and $\beta=2$.}
\label{fig2}
\end{figure*}

\begin{figure*}[ht]
\centering
\includegraphics[width=8cm]{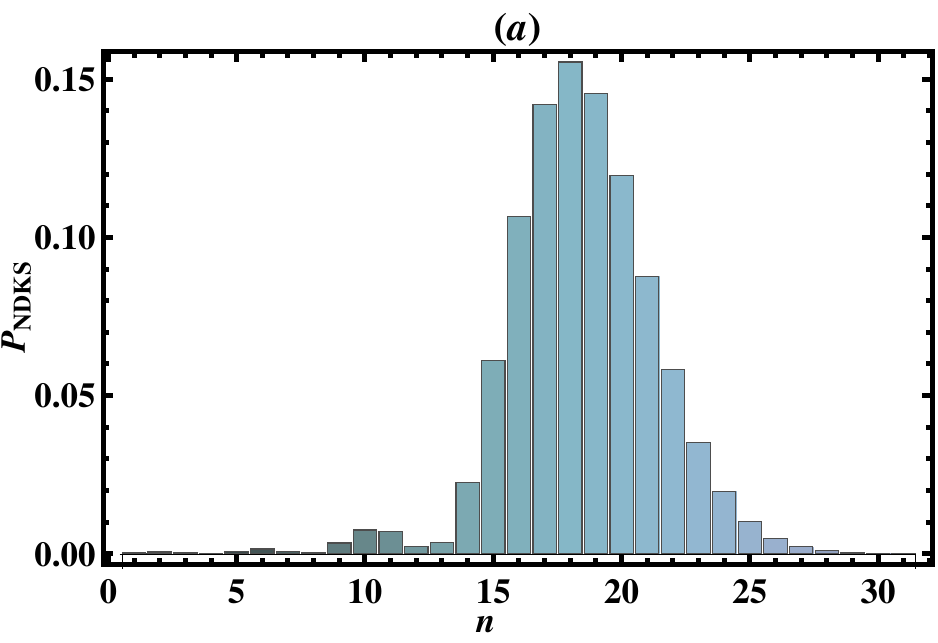}\hspace{1cm}
\includegraphics[width=8cm]{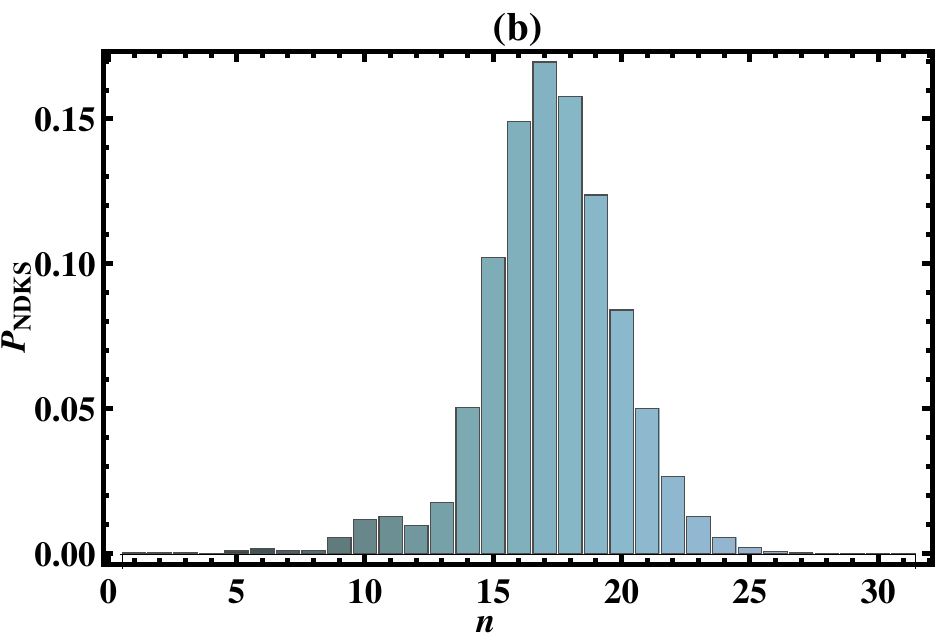}\hspace{1cm}\\
\includegraphics[width=8cm]{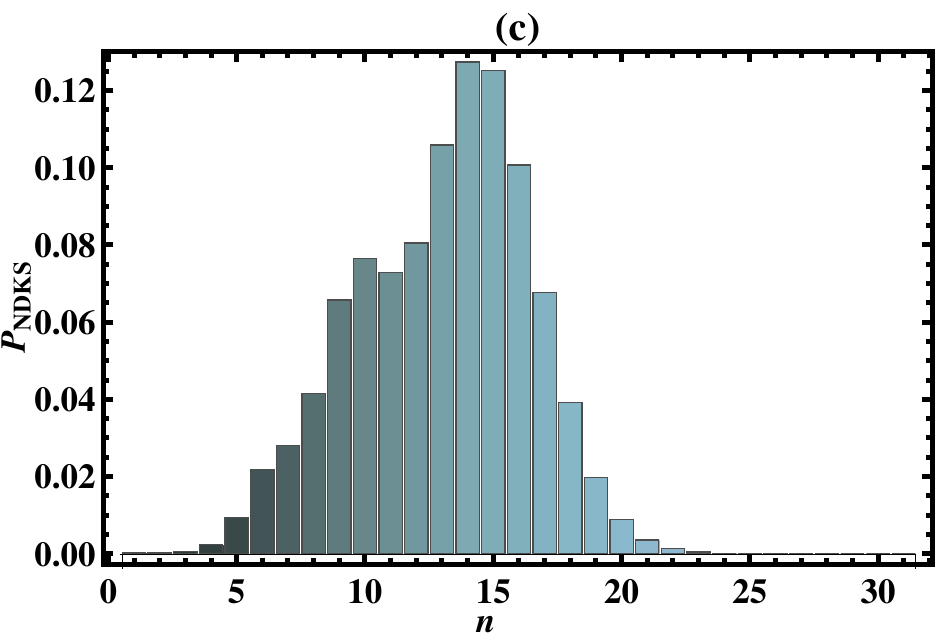}\hspace{1cm}
\includegraphics[width=8cm]{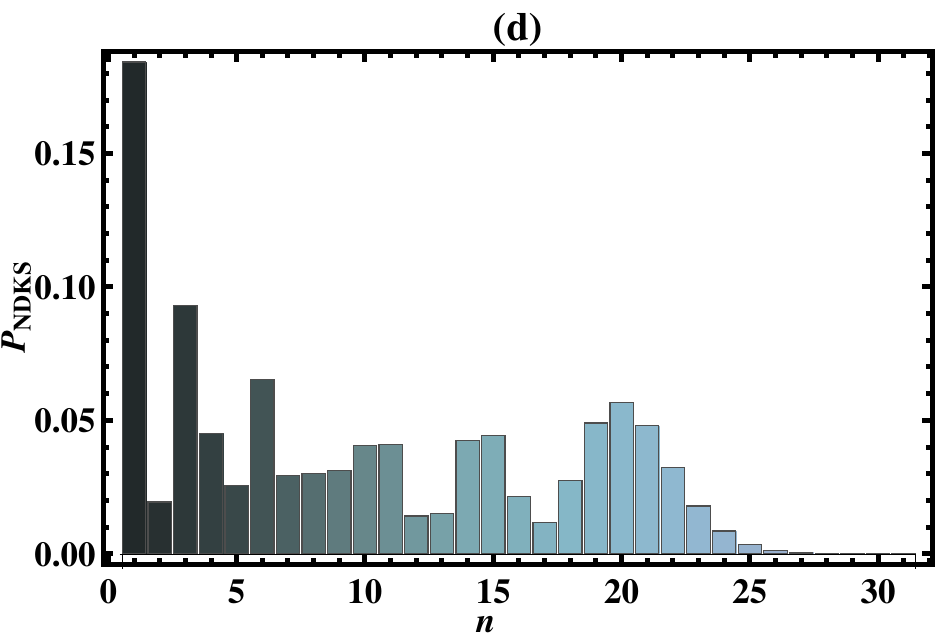}

\caption{(Color online) Photon-number distribution for the nonlinear displaced Kerr state by keeping $p$ fixed $(p=3)$ and for different $\gamma$ values: $(a)~\gamma=0.05$; $(b)~\gamma=0.12$; $(c)~ \gamma=0.25$ and $(d)~\gamma=1$. $\alpha$ and $\beta$ are taken to be real as $\alpha=1$ and $\beta=2$.}
\label{fig3}
\end{figure*}

In Fig.~\ref{fig2}, we plot $P_\mathrm{NDKS}$ versus photon number $n$ for different values of the Kerr parameter $\gamma$ that correspond to sub- and super-Poissonian statistics, which will be interpreted in the next subsection. In Fig.~\ref{fig2}(a), this distribution is for the state of maximum sub-Poissonian nature as confirmed in Fig.~\ref{fig4}. As $p=0$, this state is nothing but a usual displaced Kerr state. In Figs.~\ref{fig2}(b)-\ref{fig2}(d), the probability distributions for various $\gamma$ and $p$ are displayed. For $\gamma=0.12,~p=2$, where the statistics is sub-Poissonian, the distribution is clearly narrower than for a displaced Kerr state. However, for a higher $\gamma$ of 0.25 and $p=4$, we get a very broad distribution indicating super-Poissonian statistics in agreement with Fig.~\ref{fig4}. In Fig.~\ref{fig2}(d), $\gamma=1$ and $p=6$, $P_\mathrm{NDKS}$ becomes oscillatory and the oscillations (as a result of the interference in phase space) are much more irregular than in the case of a displaced Kerr state \cite{wilson91}. Apparently the effect of increasing $\gamma$ is to produce irregular oscillations whenever the effect of increasing $p$ below (above) the Poissonian limit is to narrow (broaden) the distribution. Figure~\ref{fig3} depicts the photon number distribution for different values of $\gamma$, with a fixed $p$. The plots resemble those in Fig.~\ref{fig2}, indicating that the shape is dependent mainly on the Kerr parameter. Fluctuations in the photon number distribution are described below.

\subsection{Mandel's $Q$ parameter}

To study the dependence on the Kerr parameter $\gamma$, we consider the Mandel's $Q$ parameter defined by \cite{mandel79}
\begin{equation}
\label{eq12}
Q = \frac{\langle(\Delta\hat{N})^2\rangle-\langle\hat{N}\rangle}{\langle\hat{N}\rangle},
\end{equation}
which measures the deviation of the photon number distribution from Poissonian statistics. Clearly $Q=0$ for a coherent field and the minimal value $Q=-1$ is obtained for Fock states, which by definition have a well-defined number of photons and for which $\langle(\Delta \hat{N})^2\rangle=0$. If $-1\leq Q<0$, the field statistics is sub-Poissonian, and the phase space distribution cannot be interpreted as a classical probability function. Therefore the negativity of $Q$ is a sufficient condition for a field to be nonclassical. Again, if $Q>0$, the states have super-Poissonian statistics, but no conclusions can be drawn about its character, e.g. the $Q$ parameter for a thermal field is always positive \cite{agarwal13}. Equation~(\ref{eq12}) requires the following quantities to be evaluated
\begin{eqnarray}\nonumber
\langle\hat{N}\rangle & = &\sum nP_\mathrm{NDKS},\\\nonumber
\langle{\hat{N}}^2\rangle & = &\sum n^2P_\mathrm{NDKS}.
\end{eqnarray}

\begin{figure}
\centering
\includegraphics[width=8cm]{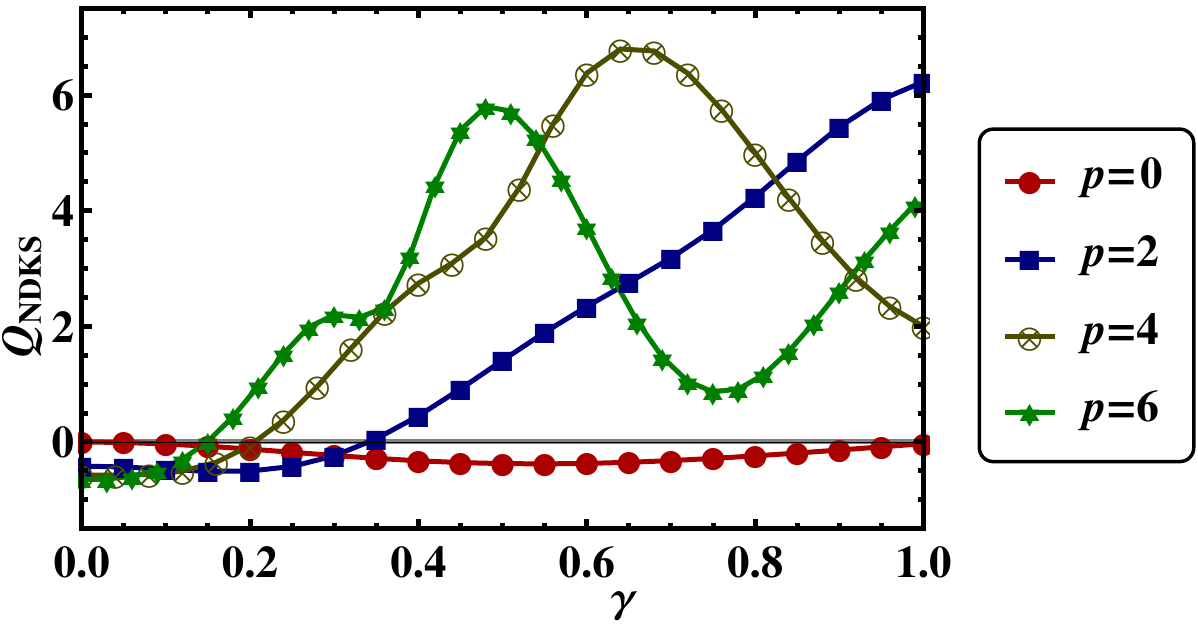}

\caption{(Color online) Mandel's $Q$ parameter as a function of the Kerr parameter $\gamma$ and for different values of the photon excitation number $p$. Other parameters are the same as in Fig.~\ref{fig2}.}
\label{fig4}
\end{figure}

In order to see the variation of the Mandel's parameter against the Kerr constant, Mandel's $Q$ is plotted as a function of $\gamma$ in Fig.~\ref{fig4}. All the $Q$ curves are partially negative and partially positive, which imply that the fields enjoying sub-Poissonian characteristic also obey super-Poissonian distribution after a certain limit of $\gamma$. For no photon excitation, $Q$ takes values less than zero, showing the sub-Poissonian character over almost the entire region of the Kerr parameter. The minimum value of $Q$ (for the case $p=0$) is $-0.5$ for $\gamma\approx 0.5$. On the contrary, the $Q$ function corresponding to $p=2$ possesses sub-Poissonian photon distribution only over a small region of weak nonlinearity ($\gamma\leq 0.22$), and this region becomes narrower while the number of photon excitations increases from $p=2$ to $p=6$. Also the Mandel's parameter for nonzero $p$ attains positive values, suggesting that the field may be of classical nature, more rapidly than the Mandel's parameter with zero $p$. Figure~\ref{fig4} clearly shows that the $Q$ parameter exhibits less nonclassicality with increasing $p$ as well as $\gamma$.

\subsection{Wigner function}
For further exploring the nonclassicality of the optical fields, the Wigner function is a powerful tool, whose partial negativity points to highly nonclassical features of quantum states \cite{louisell73,kenfack04}. In addition, the negativity is often used to present the decoherence of quantum states. For this, we consider the series form of the Wigner function in terms of the displaced number state basis as \cite{moyacessa93}
\begin{eqnarray}
W(\zeta)
= \frac{2}{\pi}\sum_{l=0}^\infty (-1)^l \langle\zeta, l|\hat{\rho}|\zeta, l\rangle,
\end{eqnarray}
where $|\zeta, l\rangle=\hat{D}(\zeta)|l\rangle$ is the displaced number state \cite{moyacessa95}.
For our nonlinear displaced Kerr state, the Wigner function is derived as
\begin{eqnarray}\nonumber
& & W_\mathrm{NDKS}(\zeta)\\
& = &
\left\{
\begin{array}{lcl}
\frac{2}{\pi}\frac{e^{-|\alpha|^2-|\beta|^2-|\zeta|^2}}{p!L_p(-|\alpha|^2)}\sum_{l=0}^\infty \frac{(-1)^l}{l!}\left|
e^{\zeta^*\beta}\sum_{m=0}^\infty\frac{\alpha^m}{m!}(m+p)!\right.\\
\left.\times e^{i\frac{\gamma}{2}(m+p)(m+p-1)}
(\beta-\zeta)^{l-m-p}\right.\\
\left.\times L_{m+p}^{l-m-p}(|\beta-\zeta|^2)\right|^2,~l\geq m+p,\\\\
\frac{2}{\pi}\frac{e^{-|\alpha|^2-|\beta|^2-|\zeta|^2}}{p!L_p(-|\alpha|^2)}\sum_{l=0}^\infty(-1)^l l!\left|
e^{\zeta^*\beta}\sum_{m=0}^\infty\frac{\alpha^m}{m!}\right.\\
\left.\times e^{i\frac{\gamma}{2}(m+p)(m+p-1)}\{-(\beta^*-\zeta^*)\}^{m+p-l}\right.\\
\left.\times L_{l}^{m+p-l}(|\beta-\zeta|^2)\right|^2,~l<m+p.
\end{array}
\right.
\end{eqnarray}

\begin{figure*}[ht]
\centering
\includegraphics[width=8cm]{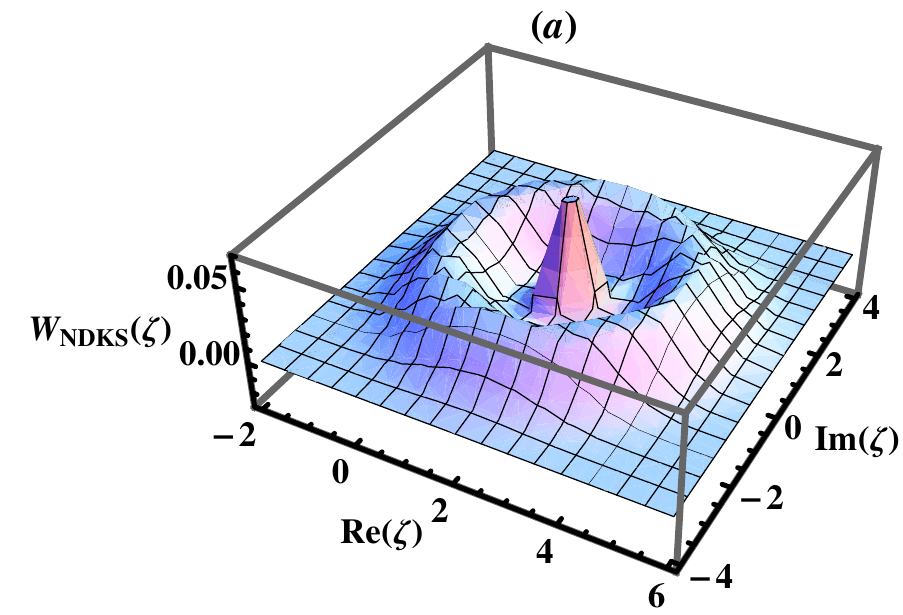}
\includegraphics[width=8cm]{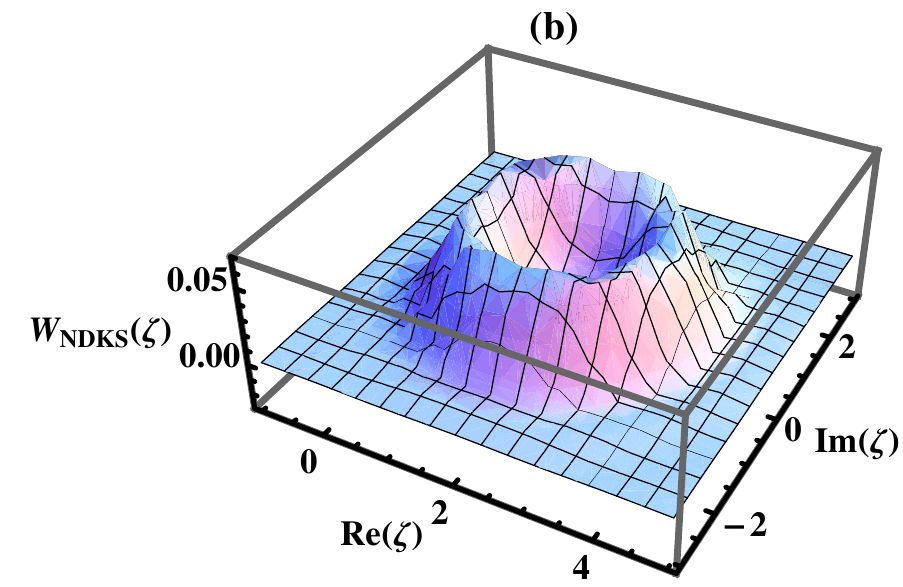}\\
\includegraphics[width=8cm]{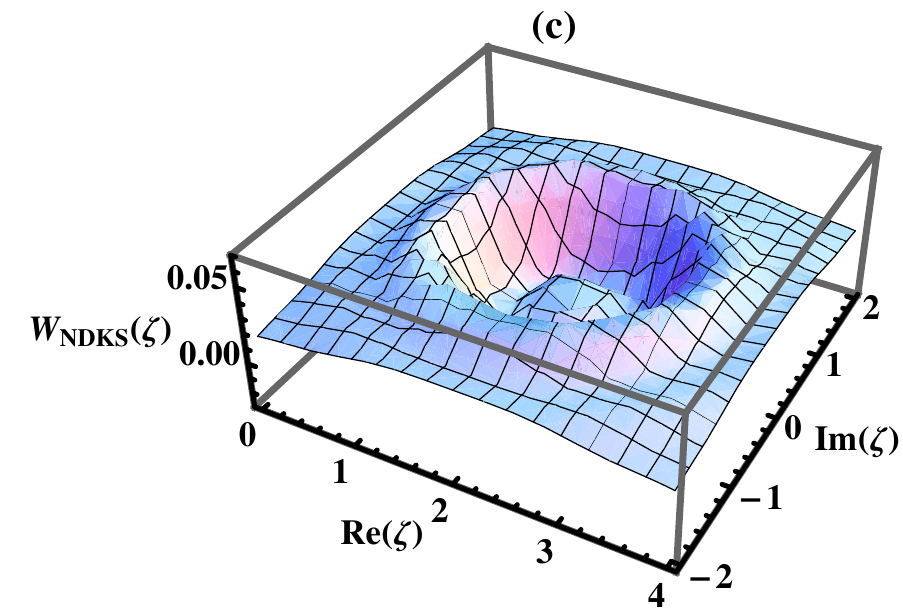}
\includegraphics[width=8cm]{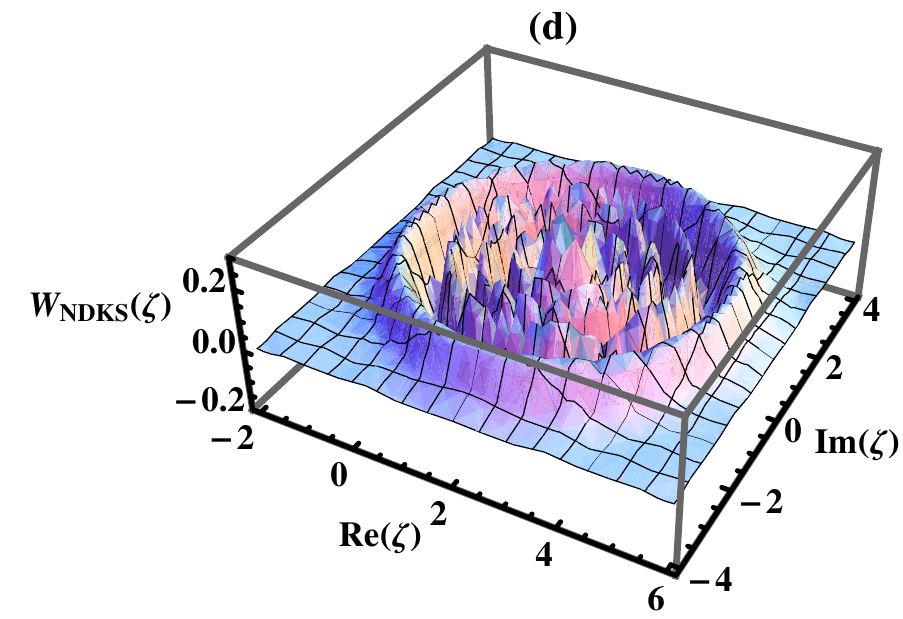}

\caption{(Color online) Wigner function $W_\mathrm{NDKS}(\zeta)$ with $\alpha=1$, $\beta=2$, and $(a)~\gamma=0.05,~p=0$; $(b)~\gamma=0.12,~p=2$; $(c)~ \gamma=0.25,~p=4$; and $(d)~\gamma=1,~p=6$.}
\label{fig5}
\end{figure*}

\begin{figure*}[ht]
\centering
\includegraphics[width=8cm]{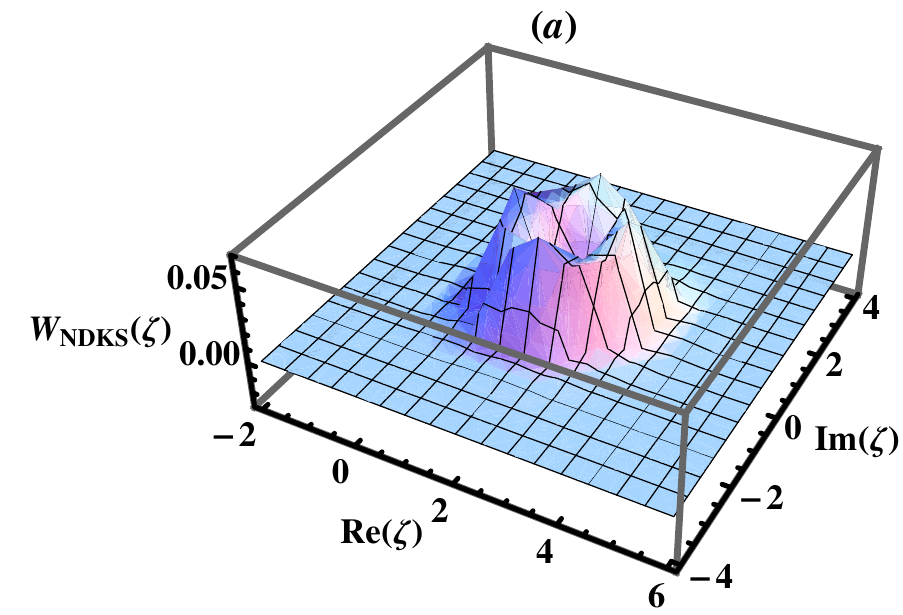}
\includegraphics[width=8cm]{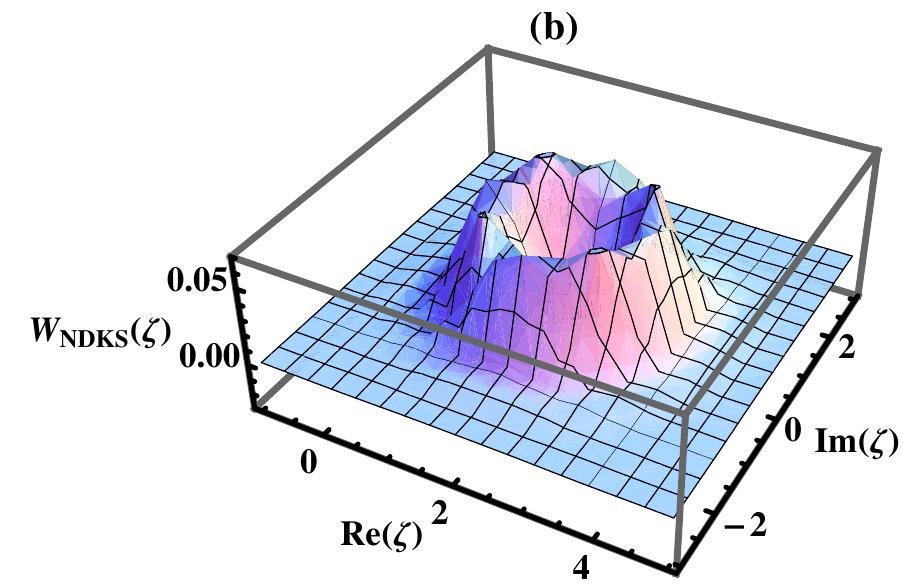}\\
\includegraphics[width=8cm]{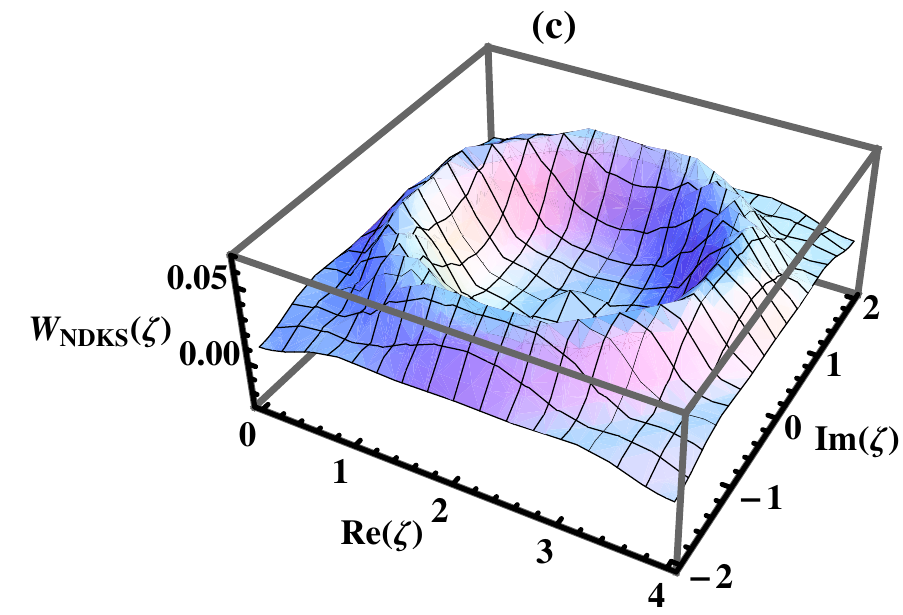}
\includegraphics[width=8cm]{fig4_wig_new.pdf}

\caption{(Color online) Wigner function $W_\mathrm{NDKS}(\zeta)$ with $\alpha=1$, $\beta=2$, $p=3$ and $(a)~\gamma=0.05$; $(b)~\gamma=0.12$; $(c)~ \gamma=0.25$ and $(d)~\gamma=1$.}
\label{fig6}
\end{figure*}

Figure~\ref{fig5} shows the behavior of the Wigner distribution in phase space for several combinations of $\gamma$ and $p$. When no photons are added ($p=0$), and for negligible $\gamma$ [Fig.~\ref{fig5}(a)], the Wigner function has a Gaussian peak surrounded by a positive dip. Here the function is positive everywhere, mimicking the classical analog. For higher values of $\gamma$ and $p=0$ [graph not shown here], we further detect that the enclosed dip becomes partially negative, pointing out the nonclassicality of the state which is quite similar to what we have found in the case of Mandel's $Q$ parameter. While raising $p$ from zero to non-zero values and increasing $\gamma$ accordingly [see Figs.~\ref{fig5}(b)-\ref{fig5}(c)], the Gaussian peak gradually transforms to a partly negative central dip. That means addition of a larger number of photons to the input coherent state in the Kerr medium leads to a partial negative region of the Wigner function. Therefore the nonlinear displaced Kerr state with nonzero $p$ is a nonclassical state in spite of the fact that the $Q$ function is super-Poissonian in nature in this domain ($\gamma=0.12$ or $0.25$). When $\gamma=1$, the Wigner function shows a very irregular pattern and a higher degree of nonclassicality. It is seen from Fig.~\ref{fig5} that increasing the photon addition number causes gain in nonclassicality of the state. This result is also verified in Fig.~\ref{fig6}, where $p$ is taken as constant at 3. While comparing the first two plots in the two figures, $p=3$ in Fig.~\ref{fig6} is greater than the $p$ values in Fig.~\ref{fig5} (a) and (b), and the Wigner functions show higher nonclassicality. But in the next two plots, the $p$ value in Fig.~\ref{fig6} is less than the values $p=4, 6$ in Fig.~\ref{fig5} (c) and (d), and the corresponding Wigner functions are less nonclassical.

\subsection{Quadrature squeezing}

In order to analyze the quantum fluctuations of the field quadratures, we consider two Hermitian operators which are combinations of photon creation and annihilation operators as
\begin{eqnarray}\nonumber
\hat{x} = \frac{\hat{a}+\hat{a}^\dag}{2},~~~~~~~~\hat{p} = \frac{\hat{a}-\hat{a}^\dag}{2i},
\end{eqnarray}
with the commutation relation $[\hat{x},~\hat{p}] = \frac{i}{2}$. They obey the Heisenberg uncertainty relation of the form
$\langle(\Delta\hat{x})^2\rangle\langle(\Delta\hat{p})^2\rangle\geq\frac{1}{16}$, and thus the quadrature squeezing occurs whenever $\langle(\Delta\hat{x})^2\rangle<\frac{1}{4}$ or $\langle(\Delta\hat{p})^2\rangle<\frac{1}{4}$. It is convenient to introduce the squeezing parameters as
\begin{eqnarray}\nonumber
s_x & = & 4\langle(\Delta\hat{x})^2\rangle-1\\\nonumber
& = & 2\langle\hat{a}^\dag\hat{a}\rangle+\langle\hat{a}^2\rangle+\langle\hat{a}^{\dag2}\rangle-
\langle\hat{a}\rangle^2-\langle\hat{a}^\dag\rangle^2-2\langle\hat{a}\rangle\langle\hat{a}^\dag\rangle,
\end{eqnarray}
and
\begin{eqnarray}\nonumber
s_p & = & 4\langle(\Delta\hat{p})^2\rangle-1\\\nonumber
& = & 2\langle\hat{a}^\dag\hat{a}\rangle-\langle\hat{a}^2\rangle-\langle\hat{a}^{\dag2}\rangle+
\langle\hat{a}\rangle^2+\langle\hat{a}^\dag\rangle^2-2\langle\hat{a}\rangle\langle\hat{a}^\dag\rangle,
\end{eqnarray}
such that the squeezing exists in $\hat{x}$ or $\hat{p}$ if $-1<s_x<0$ or $-1<s_p<0$, respectively. The expectation values
of $\hat{a}$, ${\hat{a}}^2$ and $\hat{a}^\dag\hat{a}$ are calculated as
\begin{eqnarray}\nonumber
\label{eq15}
\langle\hat{a}\rangle & = & \frac{e^{-|\alpha|^2}}{L_p(-|\alpha|^2)}\Big[\alpha(p+1)e^{ip\gamma}
{_1}F_1(2+p,2,e^{i\gamma}|\alpha|^2)\\\nonumber
& & +\beta{_1}F_1(1+p,1,|\alpha|^2)\Big],\\\nonumber
\langle{\hat{a}}^2\rangle & = & \frac{e^{-|\alpha|^2}}{L_p(-|\alpha|^2)}\Bigg[\frac{{\alpha}^2}{2}(p+2)(p+1)
e^{i(2p+1)\gamma}\\
& & \times{_1}F_1(3+p,3,e^{2i\gamma}|\alpha|^2)+2\alpha\beta(p+1)e^{ip\gamma}\\\nonumber
& & \times{_1}F_1(2+p,2,e^{i\gamma}|\alpha|^2)+{\beta}^2{_1}F_1(1+p,1,|\alpha|^2)\Bigg],\\\nonumber
\langle{\hat{a}}^\dag\hat{a}\rangle & = & \frac{e^{-|\alpha|^2}}{L_p(-|\alpha|^2)}\Big[\alpha\beta^*(p+1)
e^{ip\gamma}{_1}F_1(2+p,2,e^{i\gamma}|\alpha|^2)\\\nonumber
& & +\alpha^*\beta(p+1)e^{-ip\gamma}{_1}F_1(2+p,2,e^{-i\gamma}|\alpha|^2)\\\nonumber
& & +(p+1){_1}F_1(2+p,2,|\alpha|^2)+p{_1}F_1(1+p,1,|\alpha|^2)\\\nonumber
& & +\beta^2{_1}F_1(1+p,1,|\alpha|^2)\Big],
\end{eqnarray}
where ${_1}F_1(a;b;x)$ is the confluent hypergeometric function of the first kind. By using Eq.~(\ref{eq15}) and the relation
$\langle{\hat{a}}^\dag\rangle={\langle\hat{a}\rangle}^*$, $\langle{\hat{a}}^{\dag 2}\rangle=\langle{\hat{a}}^2\rangle^*$,
the squeezing parameters are computed.

%

\begin{figure*}[ht]
\centering
\includegraphics[width=8cm]{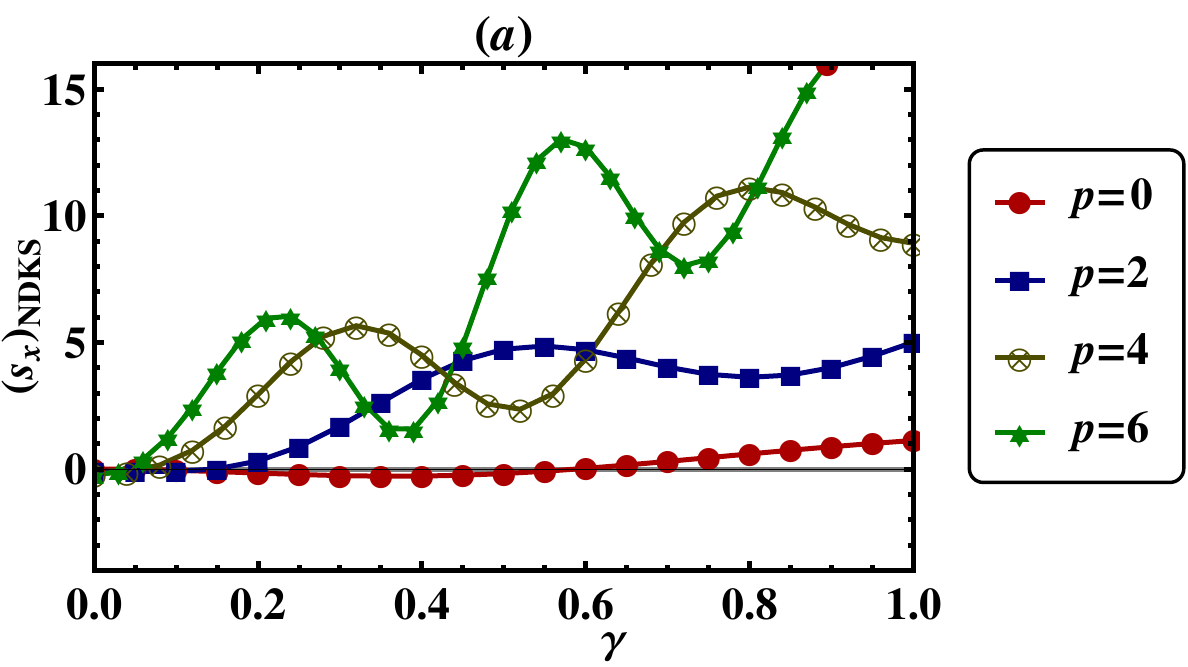}
\includegraphics[width=8cm]{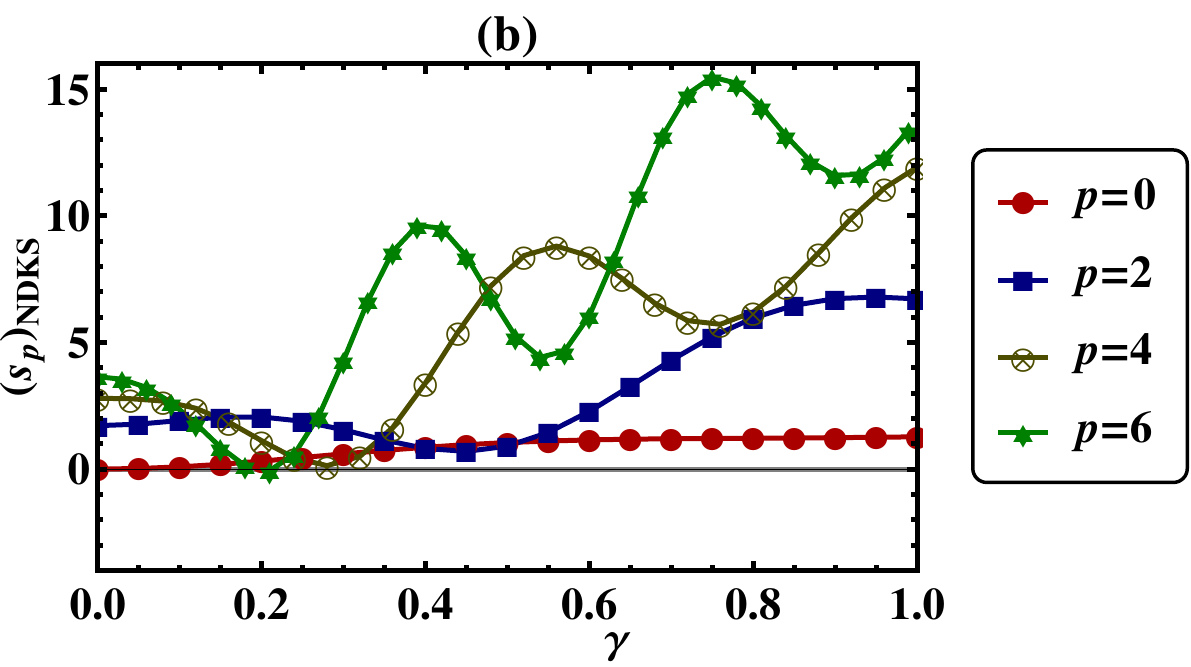}

\caption{(Color online) Squeezing parameters, $(a)~(s_x)_\mathrm{NDKS}$ and $(b)~(s_p)_\mathrm{NDKS}$ versus Kerr parameter $\gamma$ for various $p$ values.}
\label{fig7}
\end{figure*}
In order to see the influence of the photon excitation on quadrature squeezing, we have plotted the squeezing parameters against the Kerr parameter $\gamma$ in Fig.~\ref{fig7} for different values of $p$. For $p=0$, squeezing appears only in the $x$ quadrature for $0\leq\gamma\leq0.6$ with a maximum degree of quadrature squeezing of $s_x=-0.3$. For other values of $p$, no squeezing occurs in either of the quadratures. A comparison between Fig.~\ref{fig4} and Fig.~\ref{fig7} exhibits that there is a very small range of $\gamma$ where both sub-Poissonian distribution and quadrature squeezing coexist.

\subsection{Husimi-$Q$ function}

The Husimi-$Q$ function, which can be described as the coherent state expectation value of the density operator, as in Eq.~(\ref{eq5}), is the simplest quasiprobability distribution in the phase space \cite{louisell73}. This function is non-negative everywhere in the phase space, satisfying the inequality $0\leq Q(\xi)\leq\frac{1}{\pi}$, and is normalized to unity, $\int Q(\xi, \xi^*) d^2\xi = 1$. For the nonlinear displaced Kerr state $|\psi\rangle_{\mathrm{NDKS}}$, it is obtained as
\begin{eqnarray}\nonumber
& & Q_\mathrm{NDKS}(\xi)\\\nonumber & = & \frac{1}{\pi}\frac{e^{-|\alpha|^2-|\beta|^2-|\xi|^2}}{p!L_p(-|\alpha|^2)}\left|
\sum_{m=0}^\infty e^{i\frac{\gamma}{2}(m+p)(m+p-1)}e^{\xi^*\beta}\right.\\
& & \left.\times\frac{\alpha^m{(\xi^*-\beta^*)}^{m+p}}{m!}\right|^2.
\end{eqnarray}

\begin{figure*}[ht]
\centering
\includegraphics[width=6cm,height=6cm]{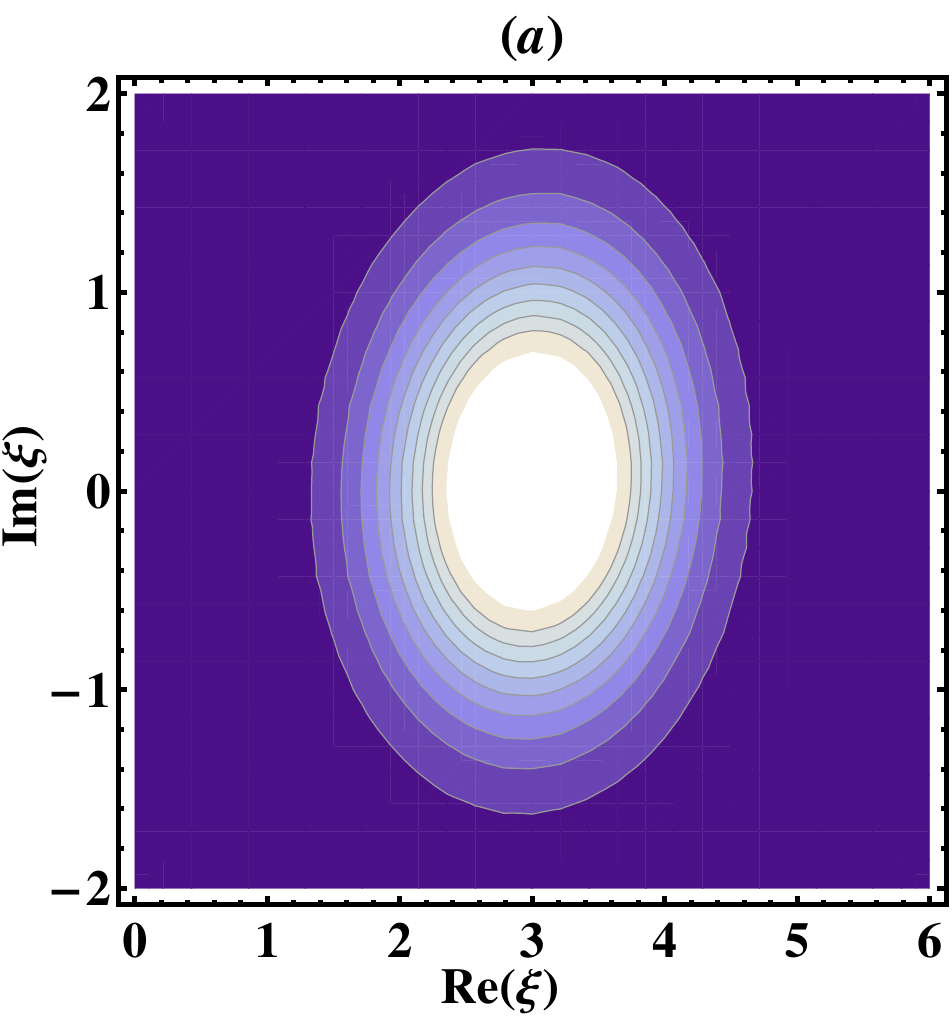}\hspace{1cm}
\includegraphics[width=6cm,height=6cm]{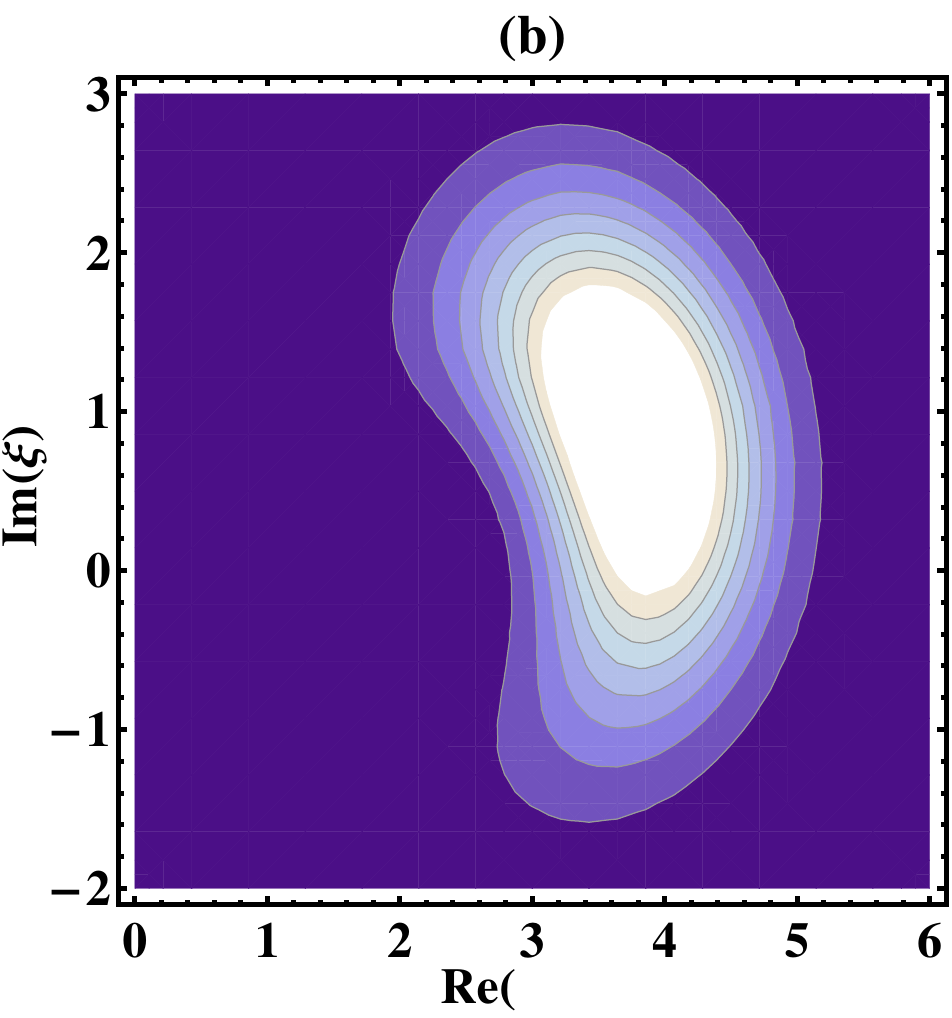}\\
\includegraphics[width=6cm,height=6cm]{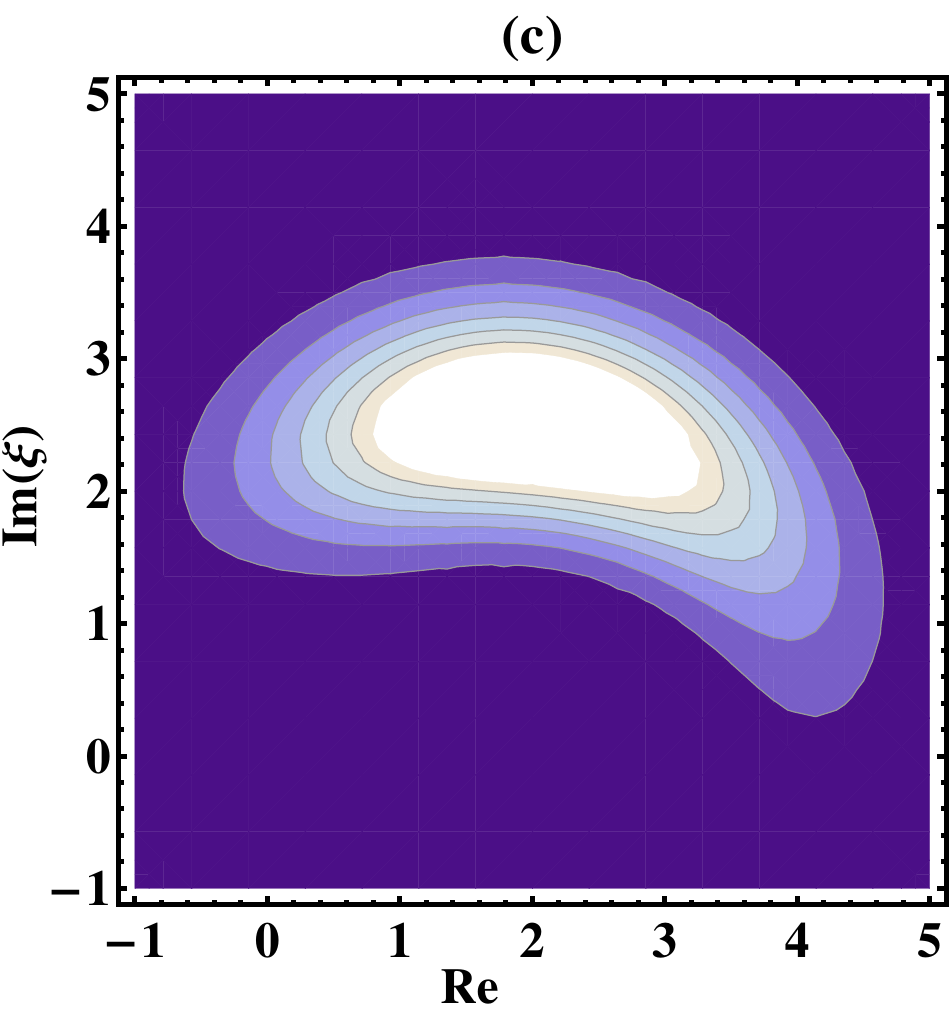}\hspace{1cm}
\includegraphics[width=6cm,height=6cm]{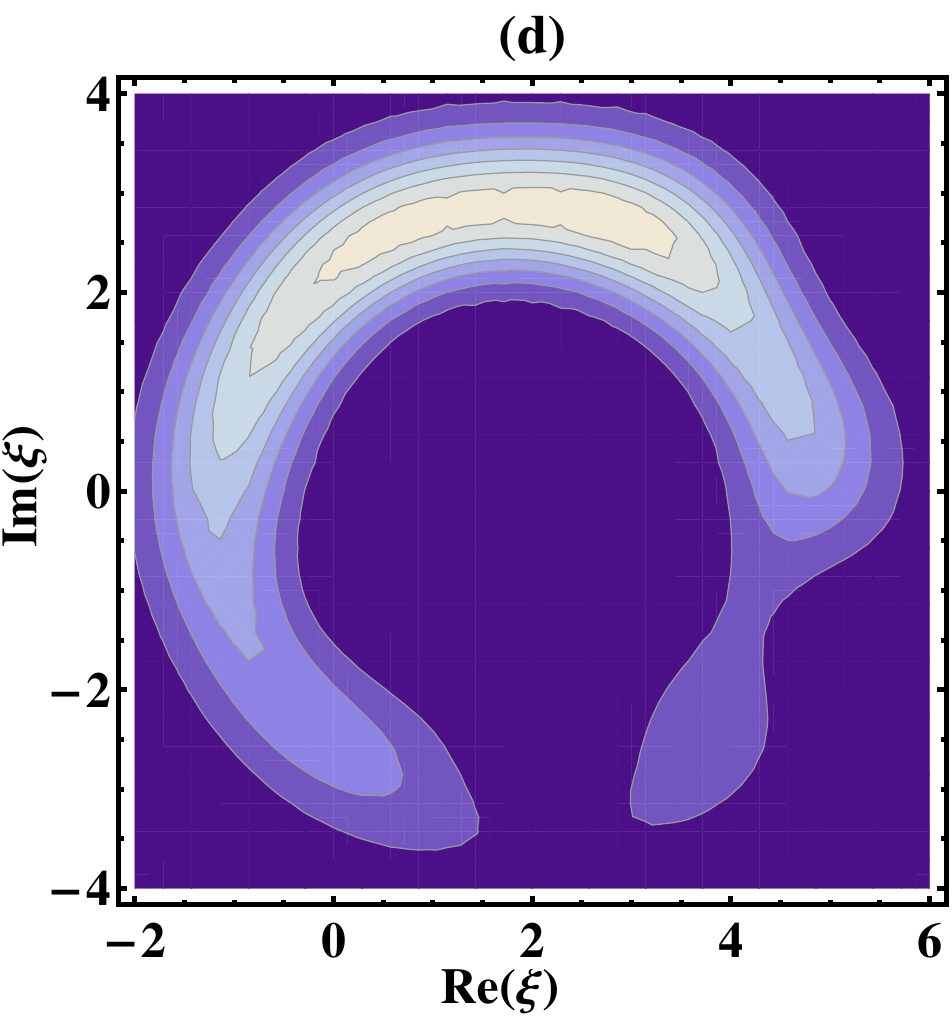}

\caption{(Color online) Contour plots of quasiprobability distribution $Q_\mathrm{NDKS}(\xi)$ as a function of $\mathrm{Re}(\xi)$ and $\mathrm{Im}(\xi)$ and for the same parameters as in Fig.~\ref{fig2}.}
\label{fig8}
\end{figure*}

In Fig.~\ref{fig8}, we present the quasiprobability distributions for the nonlinear displaced Kerr states, with the same parameters as we used to calculate the photon-number distributions in Fig.~\ref{fig2}. For $\gamma=0.05$ and $p=0$, the $Q$ function is stretched slightly, destroying its Gaussian shape as illustrated in Ref.~\cite{kitagawa86}. By comparing Fig.~\ref{fig8} with Fig.~\ref{fig4}, we observe that the quasi-probability distribution takes a banana shape, and lies in the first quadrant, corresponding to sub-Poissonian statistics with $\gamma=0.12$. For the moderately super-Poissonian state [Fig.~\ref{fig8}(c)], the shape of the distribution remains the same, and it lies in the first and second quadrants. For the highly super-Poissonian state, the phase-space distribution is ring-like whereas the photon distribution shows no regular pattern.

\subsection{Entanglement}

We now concentrate on the entanglement properties of the considered state. A single-mode nonlinear displaced Kerr state is interacted with a vacuum state $|0\rangle$ via a 50:50 beam-splitter. The resulting two-mode output state is given by
\begin{eqnarray}
|\psi\rangle_{ab} = \hat{\mathcal{B}}_{ab}\Big(|\psi_\mathrm{NDKS}\rangle_a\otimes|0\rangle_b\Big),
\end{eqnarray}
where $\hat{\mathcal{B}}_{ab}$ is the beam-splitter operator. This infinite-dimensional two-mode output state can be reduced to a discrete two-qubit system (in the $|0\rangle$, $|1\rangle$ basis) by considering a truncated nonlinear displaced Kerr state input with a low Kerr index \cite{dhar15}. Using the fact that a 50:50 beam-splitter acts on the displacement operator as $\hat{\mathcal{B}}_{ab}\hat{D}_a(\alpha)\hat{\mathcal{B}}^\dag_{ab}
= \hat{D}_a\left(\alpha/\sqrt{2}\right)\hat{D}_b\left(-\alpha/\sqrt{2}\right)$, we find
\begin{eqnarray}\nonumber
\label{eq18}
& & |\psi\rangle_{ab}\\
& = & \mathcal{A}_{ab}\Bigg[\frac{1}{\sqrt{2^{p}}}e^{i\frac{\gamma}{2}(p-1)p}\sum_{k=0}^p\binom{p}{k}^{1/2}(-1)^k|p-k\rangle_a|k\rangle_b\\\nonumber
& & +\alpha\sqrt{\frac{p+1}{2^{p+1}}}e^{i\frac{\gamma}{2}p(p+1)}\sum_{k=0}^{p+1}\binom{p+1}{k}^{1/2}(-1)^k|p+1-k\rangle_a|k\rangle_b\Bigg],
\end{eqnarray}
where $\mathcal{A}_{ab}=\frac{e^{-|\alpha|^2/2}}{[L_p(-|\alpha|^2)]^{1/2}}\hat{D}_a\left(\frac{\beta}{\sqrt{2}}\right)
\hat{D}_b\left(-\frac{\beta}{\sqrt{2}}\right)$. The density matrix for the two-mode state in Eq.~(\ref{eq18}) is $\rho_{ab}^{\mathrm{out}}=\mathcal{A}_{ab}\rho_{ab}^{\mathrm{in}}\mathcal{A}_{ab}^\dag$. Here the entanglement between the two modes of the output state is encoded into the density matrix $\rho_{ab}^{\mathrm{in}}$ since the operator $\mathcal{A}_{ab}$ acts just locally on the two modes $a$ and $b$ and thus has no effect on the entanglement properties of the output state.

If there is no photon addition to the initial coherent state, the corresponding density matrix is
\begin{equation}
\label{eq19}
\rho_{ab}^{\mathrm{in}}\Big{|}_{p=0} = \frac{1}{(1+|\alpha|^2)}\left(
\begin{array}{cccc}
1 & -\frac{\alpha^*}{\sqrt{2}} & \frac{\alpha^*}{\sqrt{2}} & 0 \\\\
-\frac{\alpha}{\sqrt{2}} & \frac{|\alpha|^2}{2} & -\frac{|\alpha|^2}{2} & 0 \\\\
\frac{\alpha}{\sqrt{2}} & -\frac{|\alpha|^2}{2} &  \frac{|\alpha|^2}{2} & 0 \\\\
0 & 0 & 0 & 0
\end{array}\right).
\end{equation}
The non-zero eigenvalues of the partial transposition matrix of $\rho_{ab}^{\mathrm{in}}\Big{|}_{p=0}$ are
\begin{eqnarray}
\left.
\begin{array}{lcl}
\lambda_{1,\,2} & = & \pm\frac{|\alpha|^2}{2(1+|\alpha|^2)},\\\\
\lambda_{3,\,4} & = & \frac{1+|\alpha|^2\pm\sqrt{1+2|\alpha|^2}}{2(1+|\alpha|^2)},
\end{array}
\right\}
\end{eqnarray}
of which only $\lambda_2$ is negative. Thus the measure of entanglement, as defined by Lee \textit{et al.} \cite{lee00}, is given by
\begin{eqnarray}
\mathrm{EP} = \log_2(1+2|\lambda_2|) = \log_2\left(\frac{1+2|\alpha|^2}{1+|\alpha|^2}\right).
\end{eqnarray}
The EP for no-photon-added NDKS is free from the Kerr parameter $\gamma$. But for $p=1$ and $p=2$, the expressions are lengthy as well as complicated, and they also depend on $\gamma$. We use the logarithmic negativity for measuring EP of the discrete-level, low-dimensional nonclassical system.

\begin{figure*}[ht]
\centering
\includegraphics[width=5.5cm]{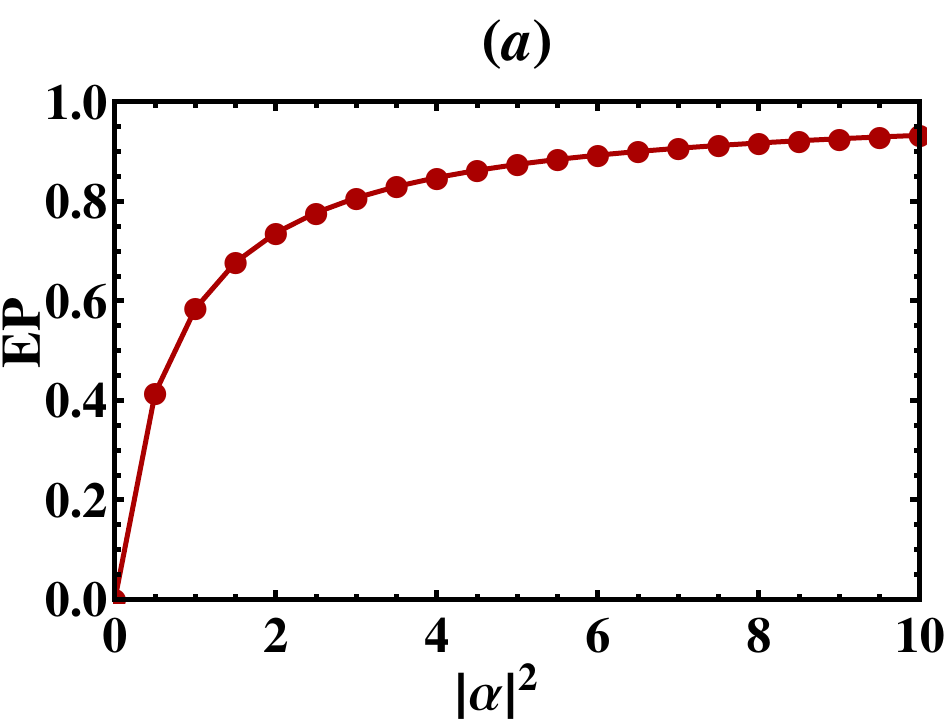}
\includegraphics[width=5.5cm]{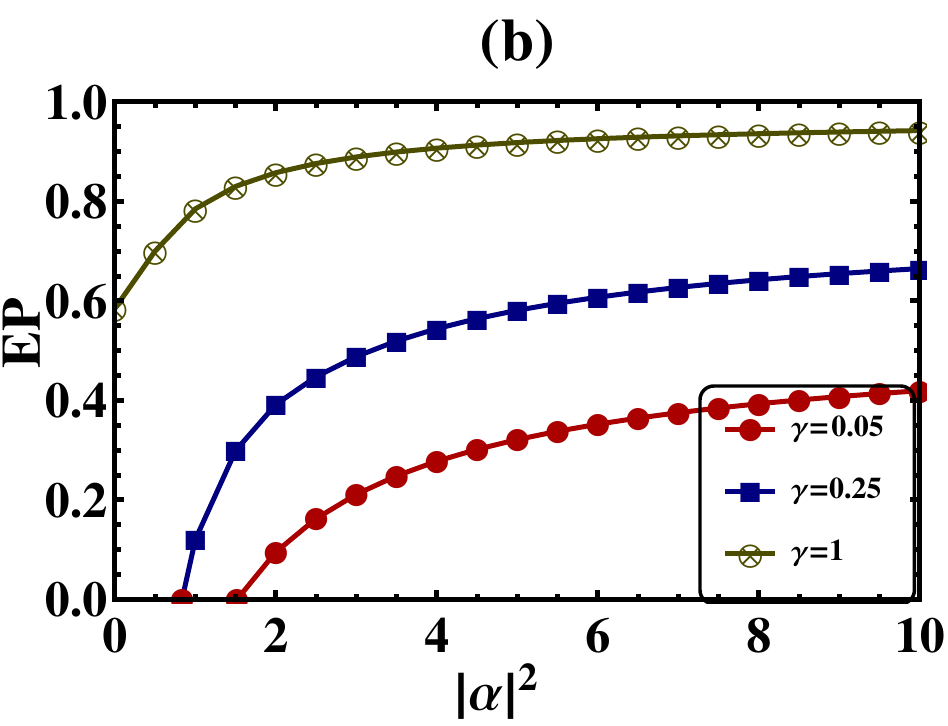}
\includegraphics[width=5.5cm]{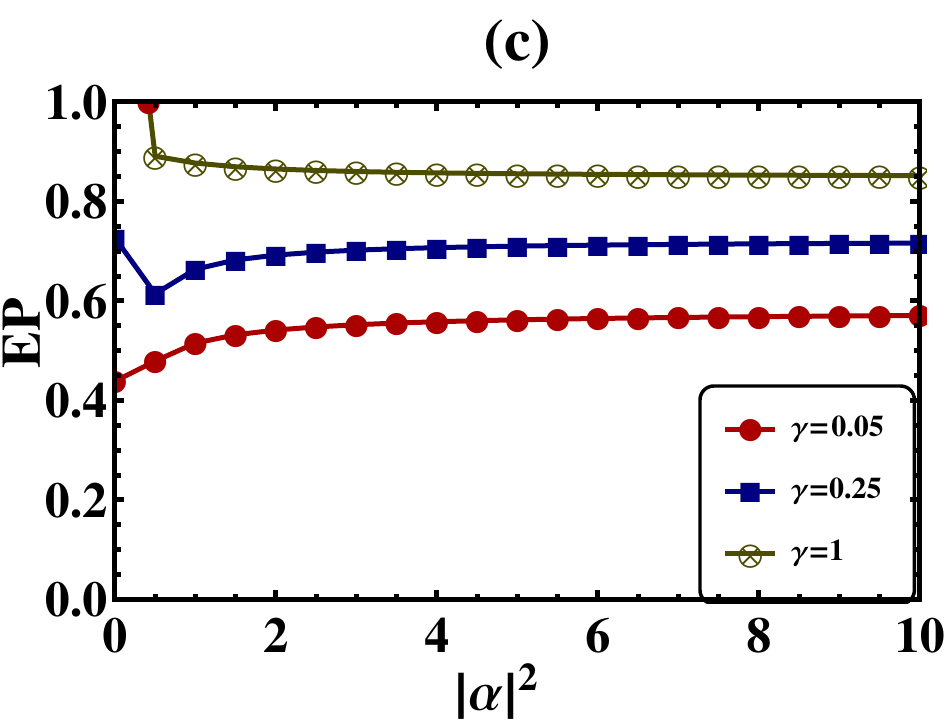}

\caption{(Color online) Entanglement potential (EP) versus coherent state amplitude $|\alpha|^2$ for two-mode output states, generated by interacting a $(a)$ no-photon-added, $(b)$ single-photon-added and $(c)$ two-photon-added displaced Kerr state with a vacuum field via a 50:50 beam-splitter. Here $\alpha$ is chosen to be real for simplicity.}
\label{fig9}
\end{figure*}

Figure~\ref{fig9} gives us the variation of EP with respect to the initial field parameter such as the amplitude of the coherent state input for different numbers of added photons. The entanglement is created here by controlling the Kerr parameter $\gamma$ so as to map the infinite-dimensional beam-splitter output state to a discrete-level optical state. We find that for a traditional displaced Kerr state (in the case of $p=0$), the EP increases rapidly while increasing $|\alpha|^2$ in the range of $0<|\alpha|^2\leq3$; it increases steadily for $3\leq|\alpha|^2\leq5$, and the maximum EP ($\approx 1$) is achieved for $|\alpha|^2\geq5$. The only point that produces no entanglement is at $|\alpha|^2=0$. The EP is independent of the Kerr index $\gamma$. Figure~\ref{fig9}(b) shows the entanglement potential of a single-photon-added displaced Kerr state against the coherent state amplitude $|\alpha|^2$. The EP remains high at most values of $|\alpha|^2$, and the maximum EP is obtained for $|\alpha|^2\geq3$. Also, the EP curve goes higher for increasing $\gamma$ from 0.05 to 1, viz. $\mathrm{EP_{max}} \approx 0.4$ for $\gamma=0.05$, $\mathrm{EP_{max}} \approx 0.6$ for $\gamma=0.25$ and $\mathrm{EP_{max}} \approx 1$ for $\gamma=1$. It means that increasing of $\gamma$ leads to more entangled states. This result is also verified by Fig.~\ref{fig9}(c). The increase of the Kerr parameter enhances the amount of entanglement over the entire range of $|\alpha|^2$. But after a certain limit ($|\alpha|^2\geq 1$), the EP is not very sensitive to changes in $|\alpha|^2$. The figure clearly shows that the EP is nonzero for almost all values of the coherent field amplitude, proving that the state remains nonclassical. This observation is in confirmation with the conclusion achieved through Wigner function analysis. As compared to the corresponding plots in Figs.~\ref{fig9}(b)-\ref{fig9}(c), we find that the EP gains higher values for higher photon-added states.

\section{Conclusion}
\label{sec4}

In this work we have presented a general method for preparing a new class of nonlinear displaced Kerr state. The nonlinear state is constructed by first sending a $p$-photon-added coherent state to a standard Kerr medium and then applying a displacement of amount $\beta$ in the phase space. We have designed an experimental set-up for possible manufacturing of the state.

The objective of the work has been to create a highly nonclassical, entangled, discrete-level, two-mode state. The infinite-dimensional interaction between the nonlinear displaced Kerr state input and the vacuum state via the beam-splitter operator is truncated to a discrete two-level system by considering a regime where the Kerr parameter, $\gamma<1$.

We have investigated the physical properties such as photon number distribution, Mandel's $Q$ parameter, Husimi-Q and Wigner functions, and quadrature squeezing for this new type of state, and then compared their distortion from the original displaced Kerr state. When the photon-addition-number $p$ is zero, the photon number plot is nothing but the usual displaced Kerr state diagram. But increasing the Kerr parameter $\gamma$ from 0.05 to 0.12 and the photon excitation number $p$ from 0 to 2 results in a clearly narrower plot than that for a displaced Kerr state plot, while increasing $\gamma$ from 0.12 to 0.25 and $p$ from 2 to 4, above the Poissonian limit, causes a wider plot. When $\gamma=1$ and $p=6$, $P_{\mathrm{NDKS}}$ becomes oscillatory, and the oscillations (as a result of the interference in phase space) are much more irregular than in the case of a displaced Kerr state. We have plotted Mandel's $Q$ parameter as a function of $\gamma$ for $\alpha=1$, $\beta=2$, $\alpha$ and $\beta$ are taken to be real. For no photon excitation, the nonlinear displaced Kerr state is sub-Poissonian over almost the entire range, with the maximum degree of sub-Poissonian statistics $Q=-0.5$ at $\gamma\approx0.5$. We have found that the $Q$ curve becomes more super-Poissonian as $p$ changes from 0 to 6, pointing out that the state may be of classical nature for nonzero $p$ values. But the Wigner distribution affirms that the state remains nonclassical for those $p$ values. When $\gamma=1$, the Wigner function is of a very peculiar pattern and shows a higher degree of nonclassicality. We have calculated the squeezing parameter as a function of $\gamma$, and observed that when $p=0$, the $x$ quadrature is squeezed only over a short range of $\gamma$, but the other quadrature is never squeezed for any value of $\gamma$. We have also found that the quadrature squeezing can co-occur with sub-Poissonian statistics for a limited range of parameters. We have described the quasiprobability distribution for the nonlinear displaced Kerr state, using the same parametric values as in Fig.~\ref{fig2}. The quasiprobability distribution of the NDKS varies from the banana-like shape corresponding to a sub-Poissonian distribution to a ring-like shape with a complicated interference structure for a highly super-Poissonian state.

Very interestingly, manipulating the field parameter, one can obtain bipartite entangled photonic states, which are useful in a wide range of applications in quantum theory, and are of fundamental importance in practical implementation of quantum teleportation theory \cite{opatrny00,kitagawa06}. The behavior of the entanglement potential is illustrated as a function of the coherent state amplitude $|\alpha|^2$. When $p=0$, the EP reaches at its maximum value of 1 for $|\alpha|^2\geq5$ and is independent of the Kerr parameter. We have also detected that $\mathrm{EP_{max}}\approx 0.4$ for $\gamma=0.05$,
$$\mathrm{EP_{max}}\approx 0.6$$ for $\gamma=0.25$, and $$\mathrm{EP_{max}}\approx 1$$ for $\gamma=1$, whenever $p=1$.
This means that the advancement of $\gamma$, keeping $p$ fixed, enhances the amount of entanglement between the two modes of the output state.

The experimental advancement in the addition of photons to an optical field ensures that our nonlinear displaced Kerr states will be put to further applications in the future. In the light of such developments the importance of theoretical studies on nonlinear photon-added states is highly significant.

\textbf{Funding Information.} Science and Engineering Research Board, Department of Science and Technology, India (SB/FTP/PS-151/2013).


\begin{thebibliography}{99}
\newcommand{\enquote}[1]{``#1''}

\bibitem{glauber63} R. J. Glauber, ``Photon correlations,'' Phys. Rev. Lett. \textbf{10}, 84-86 (1963).
\bibitem{glauber163} R. J. Glauber, ``Coherent and incoherent states of the radiation field,'' Phys. Rev. \textbf{131}, 2766-2788 (1963).
\bibitem{puri96}  R. R. Puri, and G. S. Agarwal, ``SU(1,1) coherent states defined via a minimum-uncertainty product and an
equality of quadrature variances,'' Phys. Rev. A \textbf{53}, 1786-1790 (1996).
\bibitem{agarwal91} G. S. Agarwal and K. Tara, ``Nonclassical properties of
states generated by the excitations on a coherent state,'' Phys. Rev. A \textbf{43}, 492-497 (1991).
\bibitem{tittel98} W. Tittel, G. Ribordy, and N. Gisin, ``Quantum cryptography,''
Phys. World \textbf{11}, 41-45 (1998); E. Knill, R. Laflamme and G. Milburn, ``A scheme
for efficient quantum computation with linear optics,'' Nature \textbf{409}, 46-52 (2001).
\bibitem{zavatta04} A. Zavatta, S. Viciani, and M. Bellini, ``Quantum-to-classical
transition with single-photon-added coherent states of light,'' Science \textbf{306}, 660-662 (2004).
\bibitem{hillery84} M. Hillery, R. F. O'Connell, M. O. Scully, and E. P. Wigner, ``Distribution
functions in physics: Fundamentals,'' Phys. Rep. \textbf{106}, 121-167 (1984).
\bibitem{kimble77} H. J. Kimble, M. Dagenais, and L. Mandel, ``Photon antibunching in resonance
fluorescence,'' Phys. Rev. Lett. \textbf{39}, 691-694 (1977).
\bibitem{short83} R. Short, and L. Mandel, ``Observation of sub-Poissonian photon statistics,'' Phys. Rev. Lett. \textbf{51}, 384-387 (1983).
\bibitem{dodonov02} V. V. Dodonov, `` Nonclassical states in quantum optics: a squeezed review of the first 75 years,'' J. Opt. B: Quant. Semiclass. Opt. \textbf{4}, R1-R33 (2002).
\bibitem{wenger04} J. Wenger, R. Tualle-Brouri, and P. Grangier, ``Non-Gaussian statistics
from individual pulses of squeezed light,'' Phys. Rev. Lett. \textbf{92}, 153601 (2004).
\bibitem{kim05} M. S. Kim, E. Park, P. L. Knight, and H. Jeong, ``Nonclassicality of a photon-subtracted
Gaussian field,'' Phys. Rev. A \textbf{71}, 043805 (2005).
\bibitem{lee95} C. T. Lee, ``Theorem on nonclassical states,'' Phys. Rev. A \textbf{52}, 3374-3376 (1995).
\bibitem{jones97} G. N. Jones, J. Haight, and C. T. Lee, ``Nonclassical effects in the
photon-added thermal state,'' J. Opt. B: Quant. Semiclass. Opt. \textbf{9}, 411-418 (1997).
\bibitem{filho96} R. L. Filho, and W. Vogel, ``Nonlinear coherent states,''  Phys. Rev. A \textbf{54}, 45604563 (1996).
\bibitem{manko97} V. L. Man'ko, G. Marmo, E. C. G. Sudarshan, and F. Zaccaria, ``f-oscillators and
nonlinear coherent states,'' Phys. Scr. \textbf{55}, 528-541 (1997).
\bibitem{recamier06} J. R\'{e}camier, W. L. Mochn\'{a}n, M. Gorayeb, J. L. Paz, and R. J\'{a}uregui, ``Uncertainty relations for a deformed
oscillator,'' Int. J. Modern Physics B \textbf{20}, 1851-1859 (2006).
\bibitem{roman14} R. Rom\'{a}n-Ancheyta, C. Gonz\'{a}lez Guti\'{e}rrez, and J. R\'{e}camier, ``Photon-added nonlinear coherent states for a one-mode field in a Kerr medium,'' J. Opt. Soc. Am. B \textbf{31}, 38-44 (2014).
\bibitem{shiraski90} M. Shirasaki, and H. A. Haus, ``Squeezing of pulses in a nonlinear interferometer,'' J. Opt. Soc. Am. B \textbf{7}, 30-34 (1990).
\bibitem{shiraski91} M. Shirasaki, ``Quantum-noise reduction in a phase-sensitive interferometer using nonclassical light produced through Kerr media,'' Opt. Lett. \textbf{16}, 171-173 (1991).
\bibitem{kitagawa86} M. Kitagawa, and Y. Yamamoto, ``Number-phase minimum-uncertainty state with reduced number uncertainty in
a Kerr nonlinear interferometer,''  Phys. Rev. A \textbf{34}, 3974-3988 (1986).
\bibitem{yurke86} B. Yurke, and D. Stoler, ``Generating quantum mechanical superpositions of macroscopically
distinguishable states via amplitude dispersion,'' Phys. Rev. Lett. \textbf{57}, 13-16 (1986).
\bibitem{gerry94} C. Gerry, and R. Grobe, ``Statistical properties of squeezed Kerr states,'' Phys. Rev. A \textbf{49}, 2033-2039 (1994).
\bibitem{wilson91} A. D. Wilson-Gordon, V. Buzek, and P. L. Knight, ``Statistical and phase properties of
displaced Kerr states,'' Phys. Rev. A \textbf{44}, 7647-7656 (1991).
\bibitem{zavatta08} A. Zavatta, V. Parigi, M. S. Kim, and M. Bellini, ``Subtracting photons from arbitrary light
fields: experimental test of coherent state invariance by single-photon annihilation,'' New J. Phys. \textbf{10}, 123006 (2008).
\bibitem{chatterjee12} A. Chatterjee, H. S. Dhar, and R. Ghosh, ``Nonclassical properties of states
engineered by superpositions of quantum operations on classical states,'' J. Phys. B: At. Mol. Opt. Phys.
\textbf{45}, 205501 (2012).
\bibitem{benlloch12} C. N. Benlloch, R. G. Patr\'{o}n, J. H. Shapiro, and N. J. Cerf, ``Enhancing quantum entanglement by photon addition
and subtraction,'' Phys. Rev. A \textbf{86}, 012328 (2012).
\bibitem{pinheiro13} P. Pinheiro, and R. Ramos, ``Quantum communication with photon-added coherent states,''
Quant. Inf. Process. \textbf{12}, 537-547 (2013).
\bibitem{kim08} M. S. Kim, H. Jeong, A. Zavatta, V. Parigi, and M. Bellini, ``Scheme for
proving the bosonic commutation relation using single-photon interference,'' Phys. Rev. Lett. \textbf{101}, 260401 (2008).
\bibitem{sanders92} B. C. Sanders, and G. J. Milburn, ``Quantum limits to all-optical phase shifts in a Kerr nonlinear medium,'' Phys. Rev. A
\textbf{45}, 1919-1923 (1992).
\bibitem{waks06} E. Waks, and J. Vuckovic, ``Dispersive properties and large Kerr nonlinearities using dipole-induced transparency
in a single-sided cavity,'' Phys. Rev. A \textbf{73}, 041803(R) (2006).
\bibitem{asboth05} J. K. Asboth, J. Calsamigila, and H. Ritsch, ``Computable measure of nonclassicality
for light,'' Phys. Rev. Lett. \textbf{94}, 173602 (2005).
\bibitem{paris99} M. G. A. Paris, ``Entanglement and visibility at the output of a Mach-Zehnder
interferometer,'' Phys. Rev. A \textbf{59}, 1615-1621 (1999); P. van Loock and S. L. Braunstein, ``Multipartite
entanglement for continuous variables: a quantum teleportation network,'' Phys. Rev. Lett. \textbf{84}, 3482-3485 (2000).
\bibitem{ushadevi06} A. R. Usha Devi, R. Prabhu, and M. S. Uma, ``Non-classicality of photon added coherent and thermal
radiations,'' Euro. Phys. J. D \textbf{\textbf{40}}, 133-138 (2006).
\bibitem{sudarshan63} E. C. G. Sudarshan, ``Equivalence of semiclassical and quantum mechanical descriptions
of statistical light beams,''  Phys. Rev. Lett. \textbf{10}, 277-278 (1963).
\bibitem{schliech87} W. Schleich, and J. A. Wheeler, ``Oscillations in photon distribution of squeezed states and interference in
phase space,'' Nature \textbf{326}, 574-577 (1987).
\bibitem{agarwal92} G. S. Agarwal, and K. Tara, ``Nonclassical character of states exhibiting no squeezing or sub-Poissonian
statistics,'' Phys. Rev. A \textbf{46}, 485-488 (1992).
\bibitem{mundarain04} D. F. Mundarain, and J. Stephany, ``Husimi's $Q(\alpha)$ function and quantum
interference in phase space,'' J. Phys. A: Math. Gen. \textbf{37}, 3869-3879 (2004).
\bibitem{vidal02} G. Vidal, and R. F. Werner, ``Computable measure of entanglement,'' Phys. Rev. A \textbf{65}, 032314 (2002).
\bibitem{peres96} A. Peres, ``Separability criterion for density matrices,''  Phys. Rev. Lett. \textbf{77}, 1413-1415 (1996).
\bibitem{horodecki97} M. Horodecki, P. Horodecki, and R. Horodecki, ``Inseparable two spin-$\frac{1}{2}$ density matrices can be
distilled to a singlet form,'' Phys. Rev. Lett. \textbf{78}, 574-577 (1997).
\bibitem{sivakumar99} S. Sivakumar, ``Photon-added coherent states as nonlinear coherent
states,'' J. Phys. A: Mathe. Gen. \textbf{32}, 3441-3448 (1999).
\bibitem{zavatta05} A. Zavatta, S. Viciani, and M. Bellini, ``Single-photon excitation of a coherent state: catching the elementary step
of stimulated light emission,'' Phys. Rev. A \textbf{72}, 023820 (2005).
\bibitem{tanas92} R. Tana\'{s}, B. K. Murzakhmetovs, Ts. Gantsogs, and A. V. Chizhov, ``Phase properties of displaced
number states,'' Quant. Opt. \textbf{4}, 1-7 (1992).
\bibitem{mandel79} L. Mandel, ``Sub-Poissonian photon statistics in resonance fluorescence,'' Opt. Lett. \textbf{4}, 205-207 (1979).
\bibitem{agarwal13} G. S. Agarwal, \textit{Quantum Optics} (Cambridge University Press, UK, 2013).
\bibitem{louisell73} W. H. Louisell, \textit{Quantum Statistical Properties of Radiation} (Wiley, New York, 1973).
\bibitem{kenfack04} A. Kenf\'{a}ck, and K. Zyczkowski, ``Negativity of the Wigner function as an indicator of
non-classicality,'' J. Opt. B: Quant. Semiclass. Opt. \textbf{6}, 396-404 (2004).
\bibitem{moyacessa93} H. Moya-Cessa, and P. L. Knight, ``Series representation of quantum-field quasiprobabilities,'' Phys. Rev. A
\textbf{48}, 2479-2481 (1993).
\bibitem{moyacessa95} H. Moya-Cessa, ``Generation and Properties of Superpositions of
Displaced Fock States,'' J. Mod. Opt. \textbf{42}, 1741-1754 (1995).
\bibitem{dhar15} H. S. Dhar, A. Chatterjee, and R. Ghosh, ``Generating continuous variable entangled
states for quantum teleportation using a superposition of number-conserving operations,'' J. Phys. B: At. Mol. Opt. Phys. \textbf{48}, 185502 (2015).
\bibitem{lee00} J. Lee , M. S. Kim , Y. J. Park, and S. Lee, ``Partial teleportation of entanglement in a noisy environment,''
J. Mod. Opt. \textbf{47}, 2151-2164 (2000).
\bibitem{opatrny00} T. Opatrn\'{y}, G. Kurizki, and D. G. Welsch, ``Improvement on teleportation of continuous variables by
photon subtraction via conditional measurement,'' Phys. Rev. A \textbf{61}, 032302 (2000).
\bibitem{kitagawa06} A. Kitagawa, M. Takeoka, M. Sasaki, and A. Chefles A, ``Entanglement evaluation of non-Gaussian
states generated by photon subtraction from squeezed states,'' Phys. Rev. A \textbf{73}, 042310 (2006).

\end{thebibliography}
\end{document}